\documentclass[journal]{IEEEtran}
\usepackage{booktabs} 
 \usepackage{amssymb}
\usepackage{enumitem} 
\usepackage[noadjust]{cite}
\usepackage{makecell}
\usepackage{multicol}
\usepackage{array}
\usepackage{float}
\usepackage{amsmath} 
\usepackage{amssymb}
\usepackage{amsthm}
\usepackage{multirow}
\usepackage{placeins}
\usepackage{ulem} 
\usepackage{algorithm}
\usepackage[noend]{algorithmic}
\usepackage{url}
\usepackage[caption=false]{subfig}
\usepackage{stfloats} 
\usepackage{color}
\usepackage{mathtools}
\usepackage{enumitem}

\usepackage{bbm}
\newcommand{\indicatorFunction}{\mathbbm{1}} 

\newcommand{\numRep}{Q}
\newcommand{\setRep}{\mathcal{Q}}
\newcommand{\qoe}{\mathcal{E}}

\newcommand{\video}{v}
\newcommand{\setVideo}{\mathcal{V}}
\newcommand{\numVideo}{V}
\newcommand{\chunksize}{s}

\newcommand{\chunkduration}{\tau}
\newcommand{\cacheCapacity}{S}
\newcommand{\cacheStatus}{\mathcal{S}}
\newcommand{\varcacheStatus}{x}
\newcommand{\airtime}{\theta}
\newcommand{\serveFromCache}{\phi}

\newcommand{\client}{u}
\newcommand{\setClient}{\mathcal{N}}
\newcommand{\numClient}{N}
\newcommand{\bufferCapacity}{B_{\textrm{max}}}
\newcommand{\bufferLevel}{B}

\newcommand{\bufferInSecsEstimated}{\hat{\bufferLevel}}
\newcommand{\cost}{\omega}

\newcommand{\userRate}{C}

\newcommand{\bottleneckCapacity}{\Gamma_{bh}}
\newcommand{\mappingLeveltoBitrate}{q}
\newcommand{\setRequests}{\lambda}
\newcommand{\request}{r} 
\newcommand{\requestdelivered}{\widehat{\request}}
\newcommand{\qualitydelivered}{\widehat{m}}

\newcommand{\proposalName}{EdgeDASH}
\newcommand{\buffHeuristic}{BUFF}

\newcommand{\Utility}{\mathcal{U}}

\newcommand{\weight}{\mu}

\newcommand{\toleratedQualityDiff}{\Delta\qoe} 
\newcommand{\Time}{T}
\newcommand{\queueSize}{\mathcal{D}}
\newcommand\newtexthighlight[1]{\textcolor{black}{#1}}

\begin{document}
\bstctlcite{IEEEexample:BSTcontrol}
\title{
\proposalName: Exploiting Network-Assisted Adaptive Video Streaming for Edge Caching} 
\author{
	\IEEEauthorblockN{
	    Suzan Bayhan\IEEEauthorrefmark{1},  
	    Setareh Maghsudi\IEEEauthorrefmark{2}, and
		Anatolij Zubow\IEEEauthorrefmark{2}
		}\\
	\IEEEauthorblockA{
		\IEEEauthorblockA{\IEEEauthorrefmark{1}University of Twente, The Netherlands, e-mail: s.bayhan@utwente.nl}\\
	\IEEEauthorrefmark{2}Technische Universit\"at Berlin, Germany,  e-mail: \{maghsudi, anatolij.zubow\}@tu-berlin.de}
}
\maketitle
\begin{abstract}
	While edge video caching has great potential to decrease the core network traffic as well as the users' experienced latency, 
	it is often challenging to exploit the caches in current client-driven video streaming solutions due to two key reasons. 
	First,  even those clients interested in the same content might request different quality levels as a video content is encoded into multiple qualities to match a wide range of network conditions and device capabilities. 
	%
	Second, the clients, who select the quality of the next chunk to request, are unaware of the cached content at the network edge. Hence, it becomes imperative to develop network-side solutions to exploit caching. This can also mitigate some performance issues, in particular for the scenarios in which multiple video clients compete for some bottleneck capacity. 
	In this paper, we propose a network-side control logic running at a WiFi AP 
	to facilitate the use of cached video content. 
	In particular, an AP can assign a client station a different video quality than its request, in case the alternative quality provides a better utility. This includes, for example, a function of bits delivered from the cache, video bit rate, and the buffer stalls. 
	We formulate the quality assignment problem as an optimization problem and develop several heuristics with polynomial complexity. Compared to the baseline where the clients determine the quality adaptation, our proposals, referred to as \proposalName, offer higher video quality, higher cache hits, and lower stalling ratio which are essential for user's satisfaction.  Our simulations show that \proposalName~facilitates significant cache hits and decreases the buffer stalls only by changing the client's request by one quality level. Moreover, from our analysis, we conclude that the network assistance provides significant performance improvement, especially when the clients with identical interests compete for a bottleneck link's capacity.
\end{abstract}


\section{Introduction}
\label{sec:introduction}
The ever-increasing demand for connectivity and the emergence of high-bandwidth applications have pushed network operators to seek for solutions that either increase the network capacity or improve the efficiency of resources expenditure. The \textit{edge networking} paradigm is foreseen as a solution to the aforementioned challenge, which aims at delivering content or computation from the proximity of the users, thereby decreasing the traffic in the network core.
Edge caching, in particular, suggests that the content can be stored at the periphery of the network to serve users with lower latency while decreasing network traffic, thereby 
decreasing the cost of an operator due to lower backhaul or inter-ISP traffic~\cite{mehrabi2018cache}. 
These benefits are highly desirable especially for video streaming content, which is bandwidth-hungry and \newtexthighlight{might jeopardize user's satisfaction under high latency}. While a large body of literature strongly emphasizes the role of caching in decreasing the network backhaul traffic~\cite{paschos2018role}, the challenges of caching in video streaming are vastly overlooked. Although edge caching of the video content is crucial for realizing the aforementioned goals of a network operator, 
there are several challenges due to the nature of video delivery schemes. In what follows, we briefly describe some challenges.
\begin{figure}[t]
	\centering
	\includegraphics[width=0.40\textwidth]{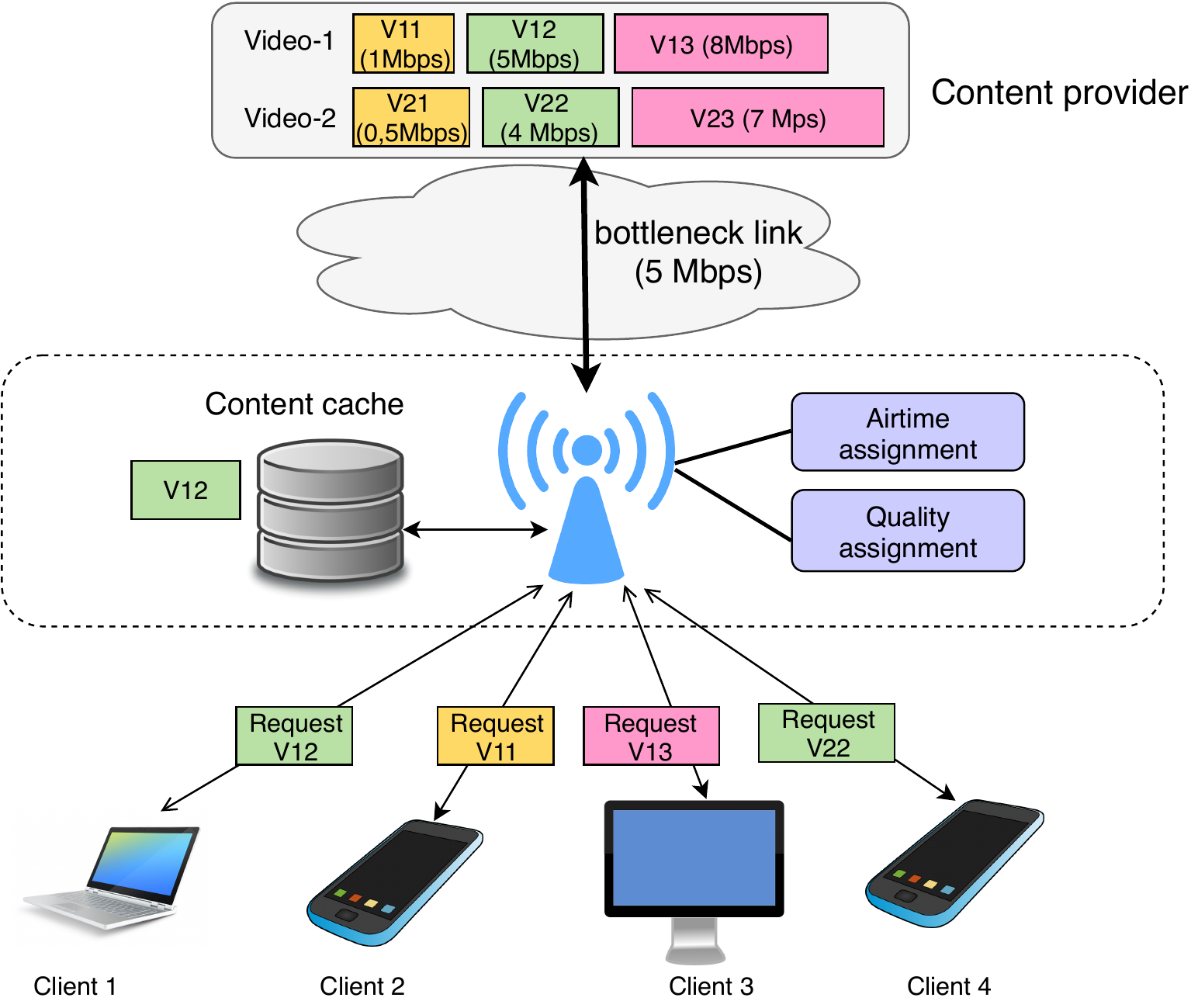}
	\caption{The WiFi AP can overwrite a request from its DASH clients to avoid congestion in its bottleneck link and to leverage its cache. In this example, the AP  delivers V12 encoded at 5 Mbps from its cache by overwriting the requests of client-2 and client-3. 
		By doing so, the AP can download the requested content V22 by client-3 with bitrate 4 Mbps from the content provider without experiencing congestion in its bottleneck link with capacity of 5 Mbps. 
	}
	\label{fig:toy-example}
	\vspace{-14pt}
\end{figure}

First, current video streaming schemes~(i.e., HTTP-based adaptive video streaming~\cite{sani2017adaptive}), rely on multi-level representation at the server-side as illustrated in Fig. \ref{fig:toy-example} and representation selection at the client-side. Although moving the control over the rate adaptation and content request process to the client offers high scalability, the content or network providers suffer from the lack of control over the delivery process. This might result in a lower caching opportunity for the network provider since the network treats different representations of the same content as different content. 
Second, \newtexthighlight{clients} are unaware of the cached content and thereby cannot favor such cached content in pure client-driven video streaming solutions. Consequently, edge cache hits for video content are limited.

In this paper, we propose to exploit the network assistance, introduced recently by MPEG server and network-assisted DASH~(SAND)~\cite{sand2018}, for realizing the potential of edge caching even if the users request different representations. While network-assistance can be implemented in all stages of the content delivery chain,
we focus here on the radio access network~(RAN), which is a WiFi network. 
We do not assume any particular rate adaptation scheme at the client. Hence, our scheme, referred to as 
\proposalName, is easy to deploy at the WiFi APs and it can work with any client player. Moreover, \proposalName~is transparent to the SAND-compliant video clients. 
Network assistance at a WiFi AP offers many benefits, including more informed decisions facilitated by the bandwidth feedback from the WiFi AP and better downlink~(DL) resource allocation at the WiFi AP considering the clients' diversity and their statistics, e.g., buffer occupancy.
\newtexthighlight{
Note that network assistance can take many forms, from airtime or quality assignment to transcoding~\cite{jin2015optimal}.
But, some network assistance functions are possible only for HTTP traffic as encrypted traffic is opaque to the intermediate network nodes~\cite{dashif2018}. In case of encrypted traffic, network assistance might require the cooperation of the video provider with the network provider to signal certain information for network assistance~\cite{petrangeli_2018}}.  

Although some research works have established the benefits of network-side solutions, e.g., \cite{nathan_minervaSigComm19, bentaleb2016sdndash, klein_nossdav17, kleinrouweler2016delivering, houdaille2012shaping}, to the best of our knowledge, there are only a few proposals, e.g., ~\cite{mehrabi2018cache}, that consider cached content delivery in a wireless RAN.   
\textit{Therefore, in this paper, our goal is to develop a quality allocation scheme at a WiFi AP to achieve high edge cache hits while taking the clients' performance into account.} To this end, our contributions are as follows.
\begin{itemize}[leftmargin=0.1in]
	\item We devise a resource allocation scheme in which the WiFi AP might overwrite the client decisions to favor (i) the consumption of the content from the edge cache and (ii) to decrease the burden on the capacity-limited bottleneck link. 
	In contrast to earlier work, e.g., \cite{mehrabi2018edge}, which transfers the quality selection decision to the network, our approach keeps the client still in the rate decision process. This design choice is motivated by  the fact that a client might prefer a certain rate due to various concerns. For example, a client with limited remaining battery or mobile data budget might prefer streaming the video at the lowest rate. 
	Moreover, due to their rich content catalogue, some users prefer video services such as YouTube to stream music~\cite{liikkanen2015music}.
	In such cases, the clients might use third party applications, e.g., FireTube, to deactivate the video, or select the lowest quality due to the above-mentioned concerns.
	If the network-side DASH solution ignores the client's decision totally, it leads to unsatisfactory user experience. 
	To remedy this, our solution defines a tolerance parameter which restricts the WiFi AP's quality assignment policy to a limited set of bitrates in the neighborhood of the quality selected by the client.
	\item Moreover, our proposal suggests that an AP allocates its DL airtime to its clients considering the assigned video qualities and the clients' statistics, e.g., buffer level, so that the clients do not experience buffer stalls or low video bitrates. Such network control is particularly useful when the core network has a bottleneck link and there are multiple 
	video  clients. 
	\item  Finally, we provide a thorough analysis of our proposals and discuss practical aspects such as their implementation using MPEG SAND protocol.
\end{itemize}

\textbf{Notation:} \newtexthighlight{Throughout the paper, we use $\client_i$ and $\video_j$ to denote client \textit{i} and video $j$, respectively. We consistently use index $i$ and index $j$ to indicate users and videos, respectively. 
Moreover, $k$ and $m$ correspondingly represent the index of a video chunk and quality level. The request of client $i$ is characterized by the following features: video $j$, chunk $k$, and quality $m$. Therefore, we denote the request as $\request_i \equiv \video_{j,k,m}$. The content that is delivered by the AP is then $\requestdelivered_i \equiv \video_{j,k,\qualitydelivered}$. 
Note that $m$ and $\qualitydelivered$ are not necessarily equal, meaning that the delivered content may have a different quality level than the requested one. We will denote all requests by $\setRequests = [\request_1,\cdots,\request_{\numClient}]$ and the content that will be delivered for these requests by $\widehat{\setRequests} = [\requestdelivered_1,\cdots,\requestdelivered_{\numClient}]$, where $\numClient$ is the number of clients.}

\section{Background on Dynamic Adaptive Video Streaming} 
\label{sec:primer}

To account for the diversity of end-user equipment and network conditions, a DASH video server divides the video into small chunks of an identical duration, e.g., 2-10 seconds long and encodes each chunk into multiple representations, e.g., bitrates. 
The number of representations\footnote{We will use the terms \textit{quality} and \textit{representation} interchangeably in the rest of the paper.} and the required bitrates are a design decision of the content provider. Chunk specification, as well as the location of each chunk, is stored in the media presentation description~(MPD) file and transmitted to the client at the beginning of a video session. Any communication between the client and the server is performed using the HTTP protocol, which is another merit of DASH: content providers can use ordinary HTTP servers, and video content can flow through network equipment without being filtered at the HTTP-friendly firewalls. 

After receiving the MPD manifest file, the client knows the properties of the video content, e.g., the number of representations and average bitrate for each representation. Based on this knowledge and some other network state information, adaptive bitrate selection~(ABR) algorithm at the client decides on which video quality to select for the next chunk. 
Usually, chunks are downloaded one by one to avoid any waste of resources if the user decides to quit the session. The downloaded chunks are stored in the client's playout buffer until their playout time comes. After some certain number of chunks are downloaded to the buffer~(e.g., a few tens of seconds), the video starts to play out. This duration between the user's request and the first playout is referred to as \textit{startup latency}.
Throughout the streaming session, there might be times where the buffer is empty resulting in video \textit{stalls}. 
User studies show that video stalls drastically decrease a user's satisfaction as well as long startup latency~\cite{seufert2015survey,petrangeli_2018,chen2015qos}.
\textit{Stalling ratio} measures how long a user stays in stall state during the video session. 
Note that initial playout policy has a significant impact on the stalling ratio, e.g., if the playout starts without a certain media in the playout buffer, the stalls are highly likely under network congestion. On the other hand, if the playout waits till many chunks are buffered to avoid stalls, the initial latency might be very high exceeding a user's patience. 

While there is no single metric capturing user's QoE commonly accepted in the literature, key factors affecting user's QoE are as follows: (i) stalling ratio, (ii) startup latency, (iii) quality switches, and (iv) visual video quality measured in terms of average bitrate of the video~\cite{chen2014qos}. Hence, an ABR scheme aims at maximizing the average bitrate while minimizing the stalling ratio and keeping the startup latency and the frequency/number of quality switches at a tolerated level to the human perception. It is a challenge to maintain the balance between these conflicting goals, especially in a multi-user setting where users compete for shared resources. 

While the literature on ABR is tremendous~\cite{sani2017adaptive}, we can categorize them into two as \textit{rate-based} or \textit{buffer-based} ABR.  In the rate-based ABR, a DASH client usually considers the rate of the recently-downloaded chunks to decide on the next chunk's quality with the aforementioned performance goals, while in the buffer-based ABR, the key parameter to consider is the client's current buffer occupancy level. There are also hybrid solutions that take both rate and buffer signals into account. However, purely-client driven ABR schemes result in inefficiency, unfairness, and instability in the multi-user DASH settings \cite{zhou_tfdash2017}. So far, several solutions are proposed to cope with the challenges mentioned above. This includes, but is not limited to: fine-tuning ABR scheme at the client~\cite{zhou_tfdash2017}, centralizing the rate adaptation decision using the SDN approach~\cite{bentaleb2016sdndash, bhat2017network}, and network-assistance for client's adaptation~\cite{mehrabi2018cache}.

Network-assistance, introduced recently by SAND~\cite{thomas2015enhancing, thomas2016applications}, aims to alleviate such performance problems by enabling protocol messages to be exchanged among network components, e.g., wireless AP, base stations, CDN edge servers. However, SAND does not specify how to efficiently use such messages, leaving space for many opportunities. In particular, SAND defines four message types: (i) status messages, (ii) metrics messages such as buffer occupancy, (iii) packets enhancing reception, 
and (iv) packets enhancing delivery. 
If a network entity is capable of processing these messages or a subset of them, then it is called a DASH-aware network element, DANE in short. Through these messages, a client and DANE can communicate for having better decisions on the next chunk to request or the next chunk to deliver. For more information, please refer to \cite{bhat2017network,thomas2015enhancing, thomas2016applications} and refer to \cite{etsi2017} for the full list of 22 SAND messages. 

\section{Related Work}
We categorize the related work into two groups, namely \textit{ network-assisted DASH} and \textit{caching for video streaming}.

\smallskip
\noindent
\textbf{Network-assisted DASH}: 
So far, several papers demonstrate the benefits of using DANEs for video streaming.  For example, in \cite{klein_nossdav17}, DANE allocates bandwidth equally among the clients and recommends a bitrate to the clients for the next chunk based on the allocated bandwidth. The client follows the recommendation of DANE only if its estimation is higher than the recommended value and the buffer exceeds a certain threshold level.  
Motivated by the shortcomings of purely client-driven rate-adaptation approaches, \cite{bentaleb2016sdndash} proposes to use an SDN controller to enable centralized control over rate-adaptation of multiple DASH clients. 
In \cite{bentaleb2016sdndash}, the SDN controller collects some information from the clients, e.g., device capabilities, buffer occupancy, and the like, to maximize QoE of each client as well as to optimize fairness and resource utilization.
Similarly, \cite{kleinrouweler2016delivering} leverages an SDN design. Both \cite{bentaleb2016sdndash} and \cite{kleinrouweler2016delivering} focus on the architecture of network-assistance and develop some schemes using SDN.
\textit{Despite sharing identical motivation with \cite{bentaleb2016sdndash}, our solution leverages SAND and retains the client-driven design of DASH. Moreover, we provide an algorithmic solution to be implemented at a WiFi AP.}

Regarding WiFi-based network assistance, \cite{kleinrouweler2016delivering} provides a comprehensive analysis of DANE assistance for rate selection and queuing on a WiFi network. \cite{zahranSAP2017} designs a stall-aware video streaming system that uses DANE messages when available. 
One of the early works on this subject is \cite{houdaille2012shaping}, where the WiFi AP applies traffic shaping to decrease the frequency of quality switching. Authors experimentally show the advantage of two video client's benefit from traffic shaping. \cite{ma_apcentric_2014} proposes to allocate AP resources using a weighted fair queuing approach and overwriting the client's decisions when necessary, i.e, the client adaptation logic is not altered.
The closest work to ours is SEBRA~\cite{khorov2017sebra}, in which a WiFi AP selects the video bitrates and the channel airtime for each video client, upon the receipt of a chunk request. SEBRA assumes a high-capacity AP to ISP link as opposed to our model with a bottleneck link. 
\newtexthighlight{With increasing number of wireless devices and video traffic, we believe that it becomes imperative to consider bottleneck links between access network and the content provider.}
Also, SEBRA solves chunk selection problem at every incoming request, whereas our proposal works only periodically, thereby attaining higher scalability.  
\textit{In addition, our solution differs from \cite{houdaille2012shaping} and \cite{khorov2017sebra} which focus only on radio access resource allocation, 
	in that we exploit edge caching in a more generic setting along with bandwidth allocation to mitigate the performance impairments due to the bottleneck links.}

\smallskip
\noindent

\smallskip
\noindent
\textbf{Caching for adaptive video streaming}: Similar to our proposal, \cite{bhat2017network} proposes to exploit network-assistance to increase the cache hits. In \cite{lee2014caching}, the authors explore the effect of caching on DASH rate adaptation algorithm. They then develop a solution to mitigate the rate fluctuations. Such fluctuations arise due to the client's overestimation of the available bandwidth when the requested chunk is cached and served directly from the cache server~\cite{lee2014caching}.
In \cite{qoeCaching2018}, the authors suggest placing the video contents on the edge servers strategically such that the initial latency remains below the maximum tolerated latency. Moreover, the clients consult the cellular base station only if unable to find the requested representation at the edge servers. 
\textit{Our work differs from \cite{qoeCaching2018} and \cite{bhat2017network} in many ways. First, we allow for delivering an approximate quality of the requested chunk if serving the latter increases the cache hits without drastically decreasing the user's satisfaction level. Second, in contrast to the earlier works that consider the cellular networks, our setting is a single WiFi cell that operates asynchronously. We discretize the continuous-time of WiFi into resource allocation intervals and quality selection intervals to mark the points of action by the WiFi AP and the DASH clients, respectively.}
A very similar study to ours is \cite{mehrabi2018cache}, which suggests maintaining some desired trade-off between the visual quality of the video and the cache hits at the ISP network. They design a coordinated bitrate selection strategy at the DASH clients such that clients will favor already-cached chunks at a slight loss of video quality to increase cache hits. While the solution of \cite{mehrabi2018cache} is for an ISP network, our solution is hosted on the WiFi radio access network which is more practical than placing the network assistance functionality deep in the core network. %

\newtexthighlight{Since end-to-end traffic encryption has become widespread, both the network assistance and the caching schemes that rely on some knowledge acquired from the  packet contents become incompatible. However, there are some works such as \cite{araldo2018caching} and \cite{gutterman2019requet} designing mechanisms to enable network control, e.g., caching, even for encrypted traffic. For example, \cite{araldo2018caching} designs a caching scheme where the content providers can leverage the benefits of caching without revealing their content to the cache provider. Another study is \cite{gutterman2019requet} which derives QoE of an encrypted video streaming session using supervised learning. }

\section{System Model}\label{sec:sysmodel}
Fig.~\ref{fig:toy-example} depicts the setting that we consider in this paper. It consists of a single WiFi BSS, e.g., a WiFi AP, and multiple WiFi stations with active video streaming sessions. The link between the WiFi AP and the serving video server is the bottleneck link with a \newtexthighlight{fixed} capacity of $\bottleneckCapacity$ Mbps.
The AP has a storage capacity of $\cacheCapacity$ bits for caching. The cache admission policy is as follows: The AP admits all the contents while it applies the least-recently-used~(LRU) replacement policy for managing its cache space. In what follows, we describe other elements of our setting. 

\textit{Video content}: 
Let $\setVideo=\{\video_1,\cdots, \video_{\numVideo}\}$ denote the set of $\numVideo$ videos.
Each video content $\video_j$ is divided into multiple chunks. Each chunk~$\video_{j,k}$ is then encoded into several representations\footnote{We use the terms quality and representation interchangeably.} denoted by $\setRep_j = \{0, 1, \cdots, \numRep_j-1\}$ with $|\setRep_j|=\numRep_j$. The representations are uniform across all chunks; consequently, we do not include $k$ in the representation description. We denote the bitrate of a representation $m$ by $\mappingLeveltoBitrate_{j,m}$ bps.
The video provider determines the duration of each video chunk, typically between 2 to 10 seconds, which may differ across different contents. We denote the chunk duration of $\video_j$ by $\chunkduration_j$ seconds. Note that \newtexthighlight{the encoding process is inherently variable, therefore} the chunks might have different sizes. As a result, the actual size of the $k^\textrm{th}$ chunk, denoted by $\chunksize_{j,k,m}$, might deviate from the average chunk size which is calculated as~$\chunksize_{j,m} = \mappingLeveltoBitrate_{j,m}\times\chunkduration_{j}$ bits~\cite{quinlan2016}. Since the MPD manifest includes only the bitrates $\mappingLeveltoBitrate_{j,m}$ to keep the file size small so that the video client can download it without a long delay, the AP knows only $\chunksize_{j,m}$, not $\chunksize_{j,k,m}$.

\textit{DASH users:} Let $\setClient=\{\client_1, \cdots, 
\client_\numClient\}$ be the set of $\numClient$ clients. 
Moreover, $\userRate_{i}$ indicates the physical layer capacity of the link connecting each client $
\client_{i}$ to the AP.
Each video client has a playout buffer of $\bufferCapacity$ seconds. Moreover, $\bufferLevel_i$ indicates the buffered video duration~(in seconds) at the client. We do not assume any particular client rate adaptation algorithm. \newtexthighlight{As described in Section \ref{sec:introduction}, we denote the requested content of client $i$ by $\request_{i}$. If required, we identify the requested content with its features, namely video $j$, chunk $k$, and quality $m$, as $\request_i \equiv \video_{j,k,m}$.} 
In case it is not necessary to specify the quality level, we omit the last index and use $\request_{i}\equiv \video_{j,k}$ to simplify the notation. In addition to the video clients, there could also be clients with background traffic. However, an AP can slice its resources for video and other less-QoS sensitive traffic~\cite{estefania_wowmom18}. Hence, we only consider the video traffic. 

Finally, we denote the cache status of the WiFi AP by $\cacheStatus=[\varcacheStatus_{j,k,m}]$, where $\varcacheStatus_{j,k,m}$ returns $1$ if chunk $k$ of video $j$ with representation $m$ is stored in the cache. As the cache capacity is limited to $\cacheCapacity$ bits, the inequality $\sum_{j}\sum_{k}\sum_{m}\varcacheStatus_{j,k,m}\chunksize_{j,k,m}\leqslant \cacheCapacity$ must hold at any time.

\begin{table}[t]
	\centering
	\caption{Key notations.}
	\begin{tabular}{|l|p{5.8cm}|} \hline 
		Notation& Description \\ \hline
		$\client_i$, $\setClient, \numClient$  &   DASH client i, set of clients, number of clients\\
		$\video_j$, $\numRep_j$, $\setVideo$ and $\numVideo$ & Video $j$, number of quality levels of video $j$, set and number of videos\\
		$\video_{j,k,m}$ & Video $j$, chunk $k$, quality level $m$ \\
		$\chunksize_{j,k,m}$& chunk size in bits for $\video_{j,k,m}$  \\
		$\chunksize_{j,m}$ & Average chunk size in bits for $\video_j$, quality level $m$ \\
		$\chunkduration_j$ & chunk duration of $\video_j$ in seconds \\
		$\bufferCapacity$, $\bufferLevel_i$ &  Buffer capacity in seconds and buffer occupancy of $\client_i$\\
		$\cacheCapacity$ &  Capacity of cache \\
		$\varcacheStatus_{j,k,m}$  & Equals 1 if $\video_j$'s chunk $k$ and quality $m$ is in cache \\
		$\airtime_i$   & Airtime allocated to $\client_i$ \\
		$\serveFromCache_i$   & Decision variable to serve $\client_i$ from cache \\
		$\userRate_i$  & Physical layer link rate of $\client_i$ \\ 
		$\bottleneckCapacity$ & Bottleneck link capacity (Mbps) \\ 
		$\mappingLeveltoBitrate_{j,m}$ & Bitrate of quality level $m$ for $\video_j$ \\ 
		$\request_i$, $\requestdelivered_i$  & Request of $\client_i$ and delivered request of $\client_i$~($\video_{j,k,\qualitydelivered}$) \\ 
		${\setRequests}$, $\widehat{\setRequests}$ & Set of all requests, and set of all delivered requests\\ 
		$\setClient^{0}$  &  Set of clients waiting for service but have already been assigned a quality level for their request. \\
		$\setClient^{1}=\setClient\setminus\setClient^0$&  Set of clients waiting to be assigned a quality level for their request.\\
		$\weight_{c}$ & Weight of cache delivery as compared to delivery from the backhaul\\
		$\toleratedQualityDiff$  &  tolerated quality difference\\  \hline
	\end{tabular}
	\label{tab:keyvars}
\end{table}

\section{\proposalName: Resource Allocation at the WiFi AP to Enable Edge Caching for Video Streaming } \label{sec:proposal}
In this section we introduce our solution, namely \proposalName, which runs on a WiFi AP for DASH- and cache-aware resource allocation. While aiming at increasing the number of cache hits, \proposalName~considers two aspects: bandwidth efficiency and QoS of the users. 
\subsection{Description of \proposalName~WiFi AP}
\begin{figure*}[t]
	\centering
	\subfloat[Quality selection intervals and AP resource allocation interval.]{\includegraphics[width=0.45\textwidth]{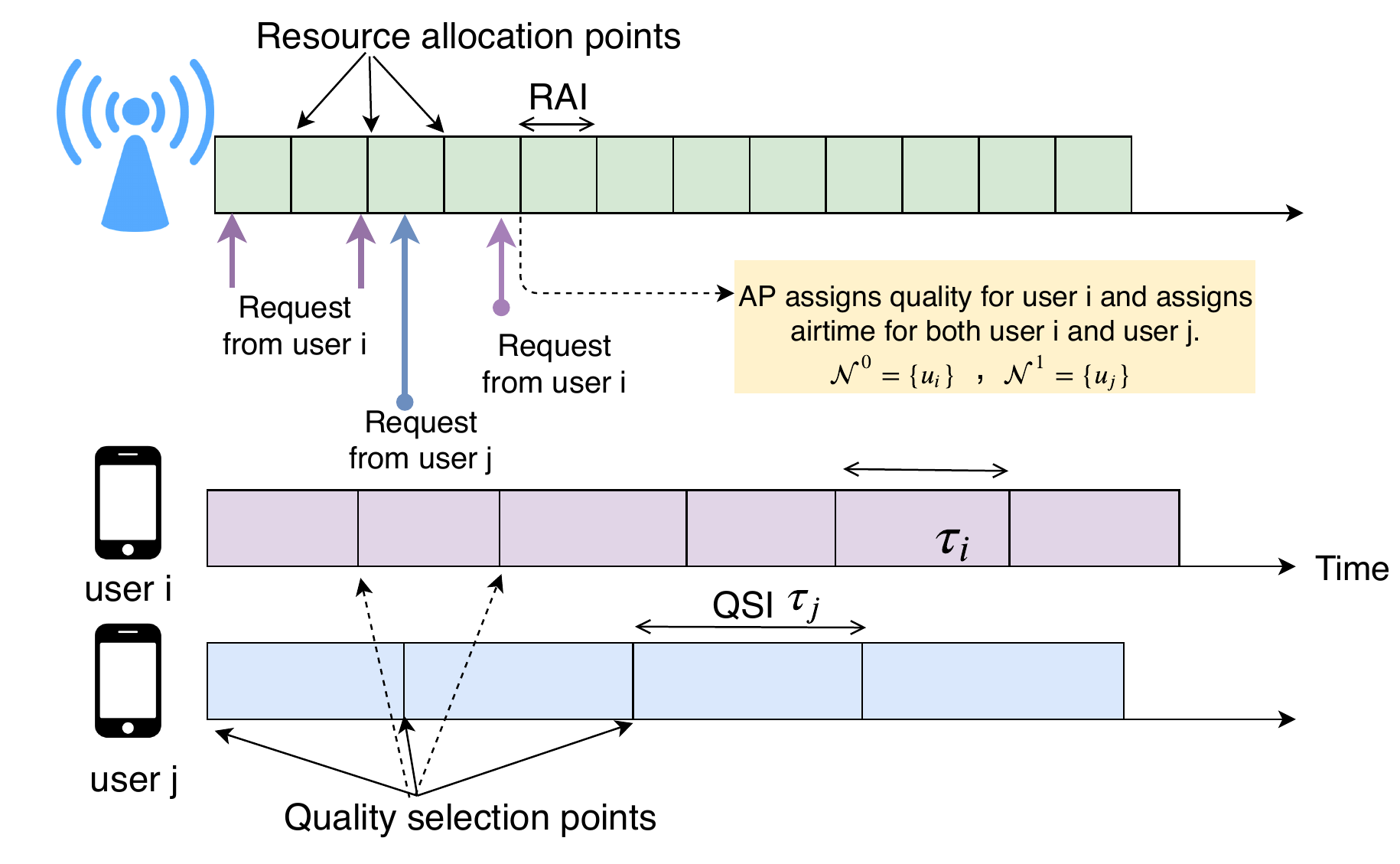}\label{fig:EdgeDASH_timeaxis}}
	\subfloat[Functional blocks in an \proposalName~WiFi AP.]{\includegraphics[width=0.42\textwidth]{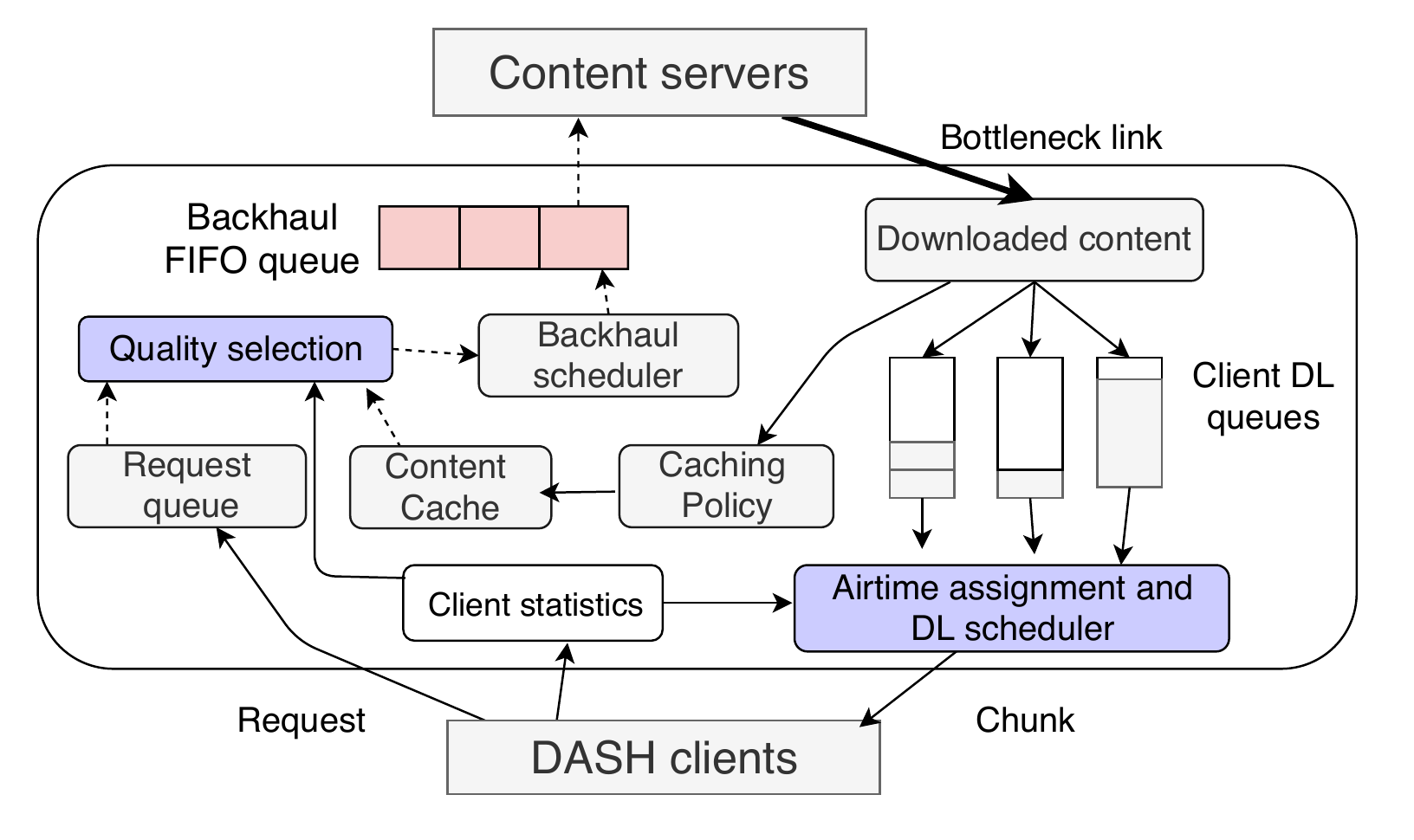}\label{fig:wifiap}} 
	\caption{(a) Resource allocation and quality selection time intervals and (b) AP functional blocks.
		\label{fig:AP}}
	\vspace{-10pt}  
\end{figure*}

Let us first explain the time scale of actions at the client's player and the WiFi AP.
Fig.~\ref{fig:EdgeDASH_timeaxis} illustrates the time points at which a client and the WiFi AP take actions. Each client decides on the next chunk's bitrate with a period approximately equal to the chunk duration of the demanded video. 
We refer to this period \textit{Quality Selection Interval~(QSI)}. 
While the QSI depends on chunk scheduling at the client player~(e.g., periodic requests, immediate 
requests after completion of each chunk, or randomized chunk scheduling~\cite{jiang2014improving}), we assume that it equals to the chunk duration~\cite{lee2014caching}.
\newtexthighlight{
This assumption stems from the steady-state dynamics of the buffer. The client requests video segments until the buffer becomes full, e.g., 10 seconds. Since the buffer cannot accommodate any new chunk, the client consumes one chunk before requesting the next chunk. Consequently, the time between two consequent chunk requests equals to chunk duration, i.e., $\chunkduration$ seconds.}
As the clients watch different videos, \newtexthighlight{the chunk duration} varies across clients.
For shorter \newtexthighlight{chunk duration, a client can react to changes swiftly in the channel or network dynamics, e.g., it selects a different quality matching the client's observed link capacity.}
As Fig.~\ref{fig:EdgeDASH_timeaxis} illustrates, the AP decides on DL resource allocation periodically, which is referred to as \textit{resource allocation interval}~(RAI).
The WiFi AP has two tasks: \textit{video quality selection} and \textit{resource allocation}. 
After collecting the client's chunk requests, the AP might overwrite the client's decision for utilizing its cache resources better. 
The WiFi AP solves the quality selection problem at the beginning of each RAI. Given that  QSI might differ across clients, the WiFi AP can select the shortest chunk duration as its RAI, \newtexthighlight{e.g.,} $T_{ap}=\min_{{\forall\video_j\in\setVideo}}(\chunkduration_j)$, where $T_{ap}$ denotes the RAI length. \newtexthighlight{However, since} the QSI is in the order of seconds whereas WiFi works in a finer time granularity for scheduling its medium access, RAI can be set in the order of milliseconds. 

\newtexthighlight{
As Fig.~\ref{fig:EdgeDASH_timeaxis} depicts, the requests arrive at the AP asynchronously due to different chunk scheduling algorithms at clients as well as different chunk duration of their consumed content. Consequently, at the beginning of a RAI, the AP needs to allocate its resources considering the new requests and those clients who have already been assigned a quality level at a previous RAI.}  
Let $\setClient^{0}$ denote the set of clients who have already been assigned quality levels. Moreover, we gather the rest of users in $\setClient^{1}$. While the AP solves the quality selection problem, it should ensure that the allocated resources are sufficient to deliver the selected quality of the video for each user. Fig.~\ref{fig:wifiap} illustrates the key functional blocks at an \proposalName~WiFi AP, as briefly described next. 
\textit{Request queue} stores the clients' chunk requests. The \textit{quality selection} module checks the request queue as well as its content cache to decide which requests to send toward the content servers and which ones to satisfy from its cache in case of any match. As the AP has a backhaul connectivity with a capacity of $\bottleneckCapacity$ (Mbps), it takes this capacity limit into account in addition to the client statistics collected from the clients.
In this step, the AP may decide to serve the user's request by a cached chunk of a different quality, if the client's streaming quality does not become deteriorated by this difference significantly. \textit{Backhaul scheduler} takes the output of the quality selection module to download requested chunks. The chunks are downloaded in a FIFO manner and in case the selected qualities exceed the capacity, there will be a certain queuing latency at the backhaul link. 
After fully downloading the chunks, the AP moves them to the DL queue of each client. The \textit{DL scheduler} then allocates the DL radio resources, i.e., its airtime, and delivers the chunks in a round-robin manner.
\subsection{Problem Formulation}

\smallskip
Recall that the request of $\client_i$ for the content $\video_{j,k,m}$ is denoted by $\request_i$. We introduce the decision variables as follows. 

\noindent\textbf{Decision variables:} The WiFi AP decides on the following three parameters for each user in $\setClient^{1}$ and only airtime for users in $\setClient^{0}$:
\begin{itemize}
	\item Quality level $\qualitydelivered_{i}\in \setRep_j$ with bitrate $\mappingLeveltoBitrate_{j,\qualitydelivered_i}$, 
	\item Share of channel airtime $\airtime_i\in[0,1]$,
	\item The client is served from the cache or not, denoted by binary variable $\serveFromCache_i\in\{0,1\}$.
\end{itemize}
Since the quality level might be modified by the AP, we will represent the assigned content as $\requestdelivered_i=\video_{j,k,\qualitydelivered_{i}}$. 

The procedure is as follows: (i) $\qualitydelivered_{i}$ takes a value from the set of available quality levels of the video that the client $i$ has requested; 
 (ii) the WiFi AP has to determine the DL airtime allocated to each user, denoted by $\airtime_i \in [0,1]$; (iii) finally, the client's request will be satisfied from the cache ~(i.e., $\serveFromCache_i=1$) or from the backhaul~(i.e., $\serveFromCache_i=0$).
We will denote the relative importance of delivery from the cache with weight $\weight_{c}$ in comparison to the delivery from the backhaul.

\smallskip
\noindent\textbf{Objective:} 
\newtexthighlight{
The objective of our proposal is to deliver a large number of bits from the cache while maintaining a high quality during the video streaming. To this end, the AP maximizes the number of delivered bits~(hence the visual quality) prioritizing the cache delivery over the backhaul delivery while considering the buffer level as video stall is known to be a  significant factor in decreasing the user satisfaction. Therefore, the AP should favor high video qualities but those that result in a buffer level above a certain threshold denoted by $\bufferLevel_{\min}$.}

Let $\mappingLeveltoBitrate_{j,\qualitydelivered_i}$ be the bitrate of the assigned quality level $\qualitydelivered_i$ of the requested video $j$. Moreover, $\bufferInSecsEstimated_i$ is the expected buffer level of a client $i$ when the current chunk with quality $\qualitydelivered_i$ is delivered to the client. In addition, we introduce $\weight_{c}$ as a tuneable parameter that reflects the desirability of cache delivery as compared to the backhaul delivery. In the next section, we describe the procedure of AP to calculate $\bufferInSecsEstimated_i$ given the assigned quality $\qualitydelivered_i$ and assigned airtime $\airtime_i$.\footnote{For the simplicity of the notation, we do not include the assigned bitrate and airtime in denoting  the estimated buffer level which depends on these two factors, i.e., $\bufferInSecsEstimated_i(\mappingLeveltoBitrate_{j,
\qualitydelivered}, \airtime_i)$.}
We first define the utility function~$\Utility_i$ as follows:
\begin{equation}
\label{eq:utility_i}
\Utility_i=
\begin{cases}
\log(\mappingLeveltoBitrate_{j,\qualitydelivered_i})(\weight_{c}\serveFromCache_i{+}(1{-}\serveFromCache_i))    +\log(\min(\bufferInSecsEstimated_i, \bufferLevel_{\max})), \\ \hspace{5cm}\text{ if }\bufferInSecsEstimated_i\geqslant \bufferLevel_{\min} \\
\log(\bufferInSecsEstimated_{i})(\weight_{c}\serveFromCache_i+(1-\serveFromCache_i)), \hspace{1cm}\text{if } \bufferLevel_{\min}>\bufferInSecsEstimated_i > 0 & \\
\bufferInSecsEstimated_{i}, \hspace{5cm}\text{otherwise.} & 
\end{cases} 
\end{equation}
The rationale behind our choice of utility function is the following: When a candidate bitrate ensures a buffer level above $\bufferLevel_{\min}$, then the associated utility is the logarithmic function of the video bitrate considering the weight of the cache delivery and the estimated buffer level. \newtexthighlight{We prefer logarithm function} to reflect the diminishing returns with increasing bitrate in terms of user's perceptual quality~\cite{nathan_minervaSigComm19}. 
When a candidate bitrate does not cause buffer stalls but cannot satisfy the target buffer level, then the corresponding utility is the logarithm of the bitrate multiplied by the cache weight. Finally, if the candidate bitrate is expected to result in negative buffer levels, we define the utility as the estimated buffer level. \newtexthighlight{Note that a negative buffer level indicates a stall with its magnitude reflecting the expected stall duration. As such, by incorporating the buffer level in the utility function, we aim at avoiding inappropriate bitrates that might result in buffer stalls. We use the expected buffer stall duration as the utility to differentiate among the quality levels that fail to sustain a smooth playout. As a result of this choice, if all quality levels are also expected to lead to buffer stalls, we select the bitrate that results in the shortest estimated buffer stall duration.} 
Under a request delivery decision \newtexthighlight{$\widehat{\setRequests}=[\requestdelivered_i], \forall u_i\in\setClient$}, and airtime allocation decision $\mathbf{\airtime}$, we formalize the objective as
\begin{equation}
\label{eq:utility}     
\underset{\qualitydelivered \in \mathcal{Q}, \airtime \in [0,1]}{\textup{maximize}} \sum_{\forall i \in \setClient^{1}}\Utility_i.    
\end{equation}
Note that in (\ref{eq:utility}), we consider the quality assignment for users in $\setClient^1$ while the performance of the users in $\setClient^0$ by introducing appropriate constraints, which we present next. 
\smallskip

\noindent\textbf{Constraints:}
\begin{itemize}[leftmargin=0.1in]
	\item If $\client_i$ is assigned a chunk with the quality level~($\qualitydelivered_i$) that exists in the cache~($x_{j,k,\qualitydelivered_i}=1$), then it is served from the cache and the content is not downloaded again. In this case, the following constraints guarantee that $\serveFromCache_i$ is 1.
	\begin{align}
	&\sum_{m\in\setRep_j}x_{j,k,m}\indicatorFunction_{(\qualitydelivered_i=m)}\leqslant  
	\serveFromCache_{i} \numRep_j\textrm{,}  ~\forall \client_i\in\setClient^{1}, \textrm{and}~\request_i \equiv \video_{j,k,m}. \nonumber\\ 
	&\sum_{m\in\setRep_j}x_{j,k,m}\geqslant 
	\serveFromCache_{i}\textrm{,}~\forall \client_i\in \setClient^{1} \textrm{},  \label{const:cacheDelivery}
	\end{align}
	where $\indicatorFunction_{g(\cdot)}$ is the indicator function that yields $1$ if the Boolean statement $g(\cdot)$ is true and $0$ otherwise. Thus, the indicator function above is 1 if the assigned quality level is $m$. 
	\item The difference in the requested and delivered quality levels is smaller than or equal to $\toleratedQualityDiff=\{0,1,2,\cdots\}$. We refer to $\toleratedQualityDiff$ as the \textit{tolerated quality difference}. We assume that the system designer selects tolerated quality difference for each client independently. Alternatively, a client ABR might also be modified to signal its tolerance level to the WiFi AP. Obviously, $\toleratedQualityDiff=0$ implies that only requested quality and no alternative is acceptable. Formally,
	\begin{align}
	\qualitydelivered_i - m  \leqslant |\toleratedQualityDiff|\textrm{,}~\forall i\in\setClient^{1} \textrm{ and } \request_i \equiv \video_{j,k,m}. \label{const:toleratedDifference}
	\end{align}

	\item As the backhaul capacity is limited to $\bottleneckCapacity$ Mbps, we have: 
	\begin{align}
	\sum_{\forall i\in\setClient^1} (1-\serveFromCache_{i}) \mappingLeveltoBitrate_{j,
\qualitydelivered_{i}} 
	\leqslant \bottleneckCapacity, \label{const:totalBackhaulCapacity}
	\end{align}
	in which we assume that downloading of the previously requested content for users in $\setClient^{0}$ is already completed. In case it is not, the AP considers the remaining bandwidth for these new requests. 
	\item The DL airtime of the WiFi AP is limited to 1, which can be formalized as:
	\begin{align}
	\sum_{\forall i\in\setClient} \airtime_i \leqslant 1.  \label{const:totalAirtimeEquals1}
	\end{align}
	For taking the uplink traffic into account, one can further restrict the allocated time, implying that $\sum_{\forall i\in\setClient} \airtime_i \leqslant \airtime$ where $0<\airtime<1$~\cite{khorov2017sebra}.
	\item As buffer stalls result in unpleasant user experience, the AP should consider the estimated buffer level of all its clients. 
	Estimated buffer level depends on the assigned quality level as well as the client's current buffer level and the client's  link capacity. 
	The link capacity is proportional to the client's airtime and its channel capacity. 
	%
	If the client's buffer already has sufficient media~(i.e., high buffer occupancy), \newtexthighlight{
the AP can select a quality representation whose bitrate is higher than the client's effective link
capacity while ensuring that the expected buffer size is nonnegative. Formally,}
	%
	\begin{align}
	\bufferInSecsEstimated_i\geqslant 0 \textrm{,} ~\forall i\in\setClient.  \label{const:buffEstimatedNonZero}
	\end{align}
\end{itemize}

Based on the discussion above, the challenge is to solve the following optimization problem. 
\begin{align} 
\max_{\qualitydelivered,\airtime,\serveFromCache}& \sum_{\forall i\in\setClient^{1}} \Utility_i\label{obj:utility_all_in_1} \\
& \text{s.t.}~(\ref{const:cacheDelivery}),  (\ref{const:toleratedDifference}), (\ref{const:totalBackhaulCapacity}),(\ref{const:totalAirtimeEquals1}), (\ref{const:buffEstimatedNonZero}), \\
&\qualitydelivered_{i} \in \setRep_j\textrm{,}~\forall i\in \setClient^{1} \textrm{ and } \request_i \equiv \video_{j,k,m} \label{const:reqDeliveredInSet}\\
&\airtime_i \in [0,1] \textrm{,}~\forall i\in \setClient  \label{const:airtimeVar} \\
& \serveFromCache_i \in \{0,1\} \textrm{,}~\forall i\in \setClient^{1}. \label{const:vartype_serve}
\end{align}
Due to the existence of decision variables belonging to a discrete set, the problem in (\ref{obj:utility_all_in_1})-(\ref{const:vartype_serve}) is computationally hard. In the next section, we address this challenge by first modeling our problem as a multiple choice knapsack problem~(MCKP), which is NP-hard~\cite{pisinger1995minimal,bednarczuk2018multi}. Afterward, we propose a heuristic soluation based on an approximation algorithm for solving MCKP~\cite{shojaei2013}. 
In our approach, we first assume equal airtimes for each client and concentrate on the quality assignment. Afterward, we assign airtimes given the quality levels. 
\section{Video Quality Assignment as a Multiple Choice Knapsack Problem} 
\label{sec:MCKP}
To simplify the NP-hard problem formulated in Section \ref{sec:proposal}, we divide it into two problems, namely (i) quality selection and (ii) airtime assignment. In brief, the solution is as follows: In the first step, we adapt an existing solution for 0-1 MCKP, namely \textit{Compositional Pareto-algebraic Heuristic}~(CPH), to our problem assuming equal airtimes for all clients. In the second step, we propose an airtime assignment approach that aims at minimizing the buffer stalls.

\subsection{Compositional Pareto-algebraic Heuristic~(CPH)}

\begin{figure*}
	\centering
	\subfloat[CPH in original form.] {\includegraphics[scale=0.45]{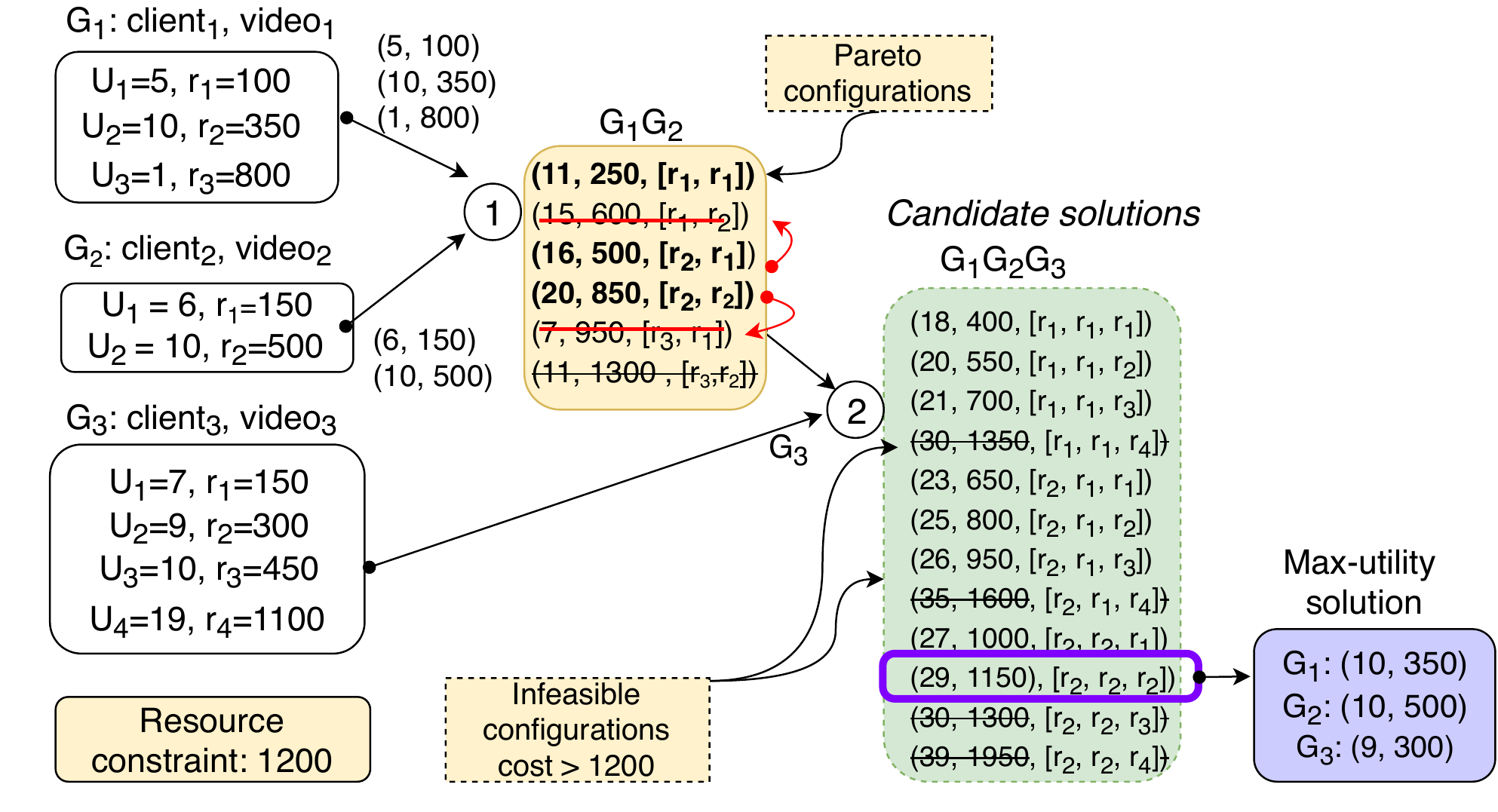}\label{fig:cph}}
	\subfloat[CPH adapted for the quality selection problem.]{\includegraphics[scale=0.45]{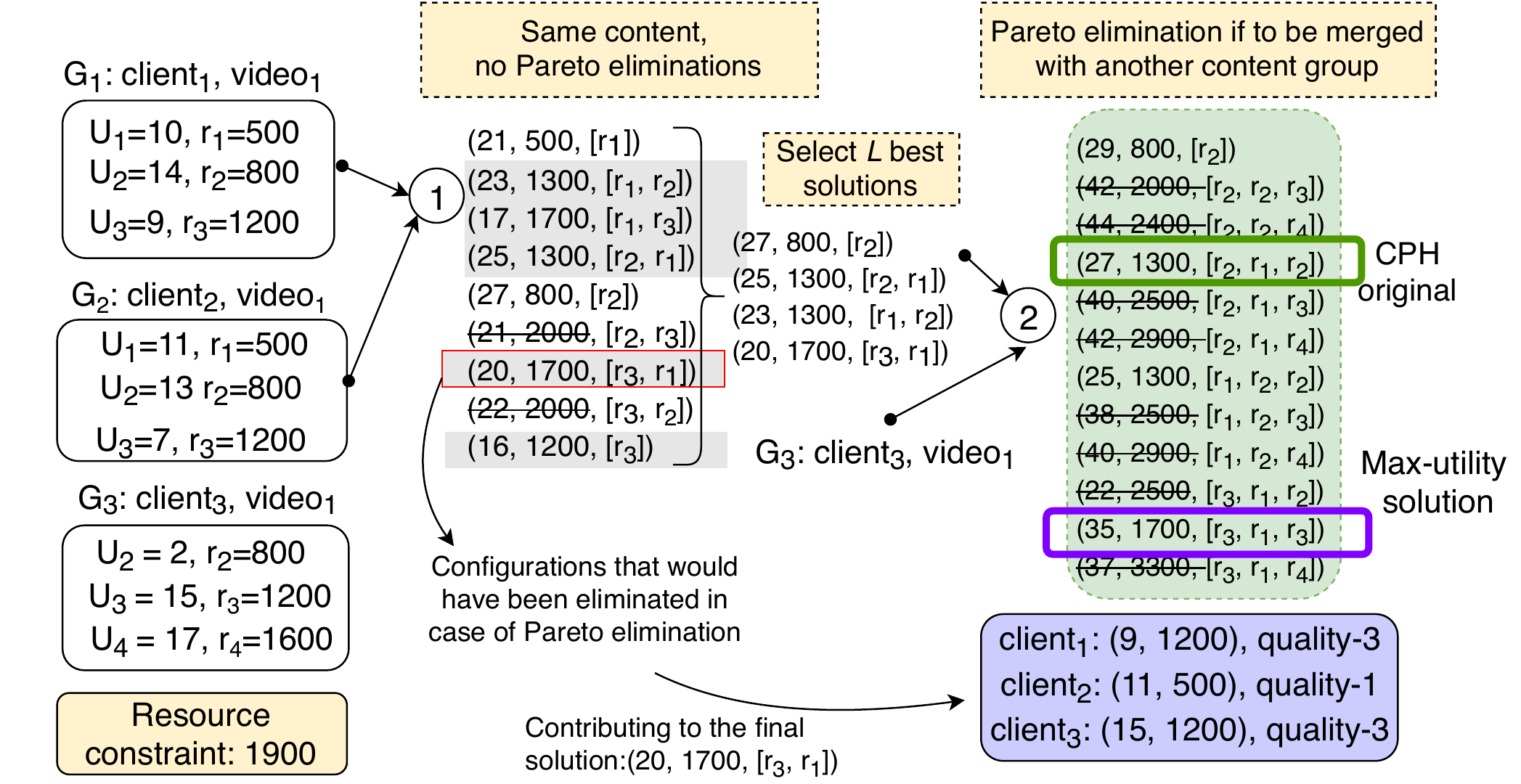}\label{fig:cph-modified}}
	\caption{
	\newtexthighlight{
		(a) A schematic overview of CPH, as explained in Section \ref{sec:MCKP}: 
		To reduce the size of each partial configuration, at each step, CPH removes configurations that are Pareto-dominated or violate the resource constraints. 
		(b) CPH adapted to our quality selection problem: The clients that request the same content are merged without Pareto elimination. This variant of CPH keeps the shaded configurations in Step 1 due to the following reason. These configurations might later yield the maximum utility when another client requests the same content. 
		In this example, the maximum utility consists of a partial configuration~(20, 1700) that would be eliminated by CPH resulting in a lower utility, i.e., 27 instead of 35. }
		\label{fig:cph-all}
		} 
\end{figure*}
In the following, we provide a brief overview of \textit{compositional Pareto-algebraic Heuristic}~(CPH)~\cite{shojaei2013}. Moreover, we propose a procedure to adapt it to solve the formulated video quality assignment problem. 

CPH is designed to solve multi-dimensional MCKP. Fig. \ref{fig:cph} shows a toy example for a single dimensional MCKP in which there are three groups~(corresponding to the clients), each with multiple items~(corresponding to the quality levels). 
For each item in each group, there is an associated utility and resource consumption value. Moreover, the knapsack has a maximum capacity (corresponding to the capacity of the bottleneck link, e.g., minimum of DL and backhaul capacity). The objective of CPH is to select an item from each group such that the total utility is maximized without violating the resource constraint in each dimension.
\newtexthighlight{Note that although CPH is designed for multidimensional MCKP problems, it does not entail extra complexity or overhead when applied to solve the single-dimensional problems.}
\begin{algorithm}[t]
	\textbf{\caption{CPH-based quality assignment~(CPH)}\label{alg:cph}}
	\begin{algorithmic} [1]
		\STATE \textbf{Input}: Requests to be assigned a quality~($\setRequests =[\request_i]$), client-AP link capacity considering equal airtime allocation, AP DL client queues~($\queueSize_i$), client buffer level~($\bufferLevel_i$), available backhaul capacity~($\bottleneckCapacity$), cache status~($\cacheStatus=[\varcacheStatus_{j,k,m}]$)
		\STATE \textbf{Output}: $\widehat{\setRequests}$: video chunks to deliver for each $\client_i$.
		\STATE Find set of tolerated quality levels~($\mathcal{M}_i$) for each client request using tolerated quality difference $\toleratedQualityDiff$.
		\STATE Calculate utility $\Utility_{i,m}$ for each  client and quality $m\in\mathcal{M}_i$. 
		\STATE Calculate the cost of delivering each quality as: \label{alg:cph:cost} $\cost_{j,m}=\mappingLeveltoBitrate_{j,m}$ if $\varcacheStatus_{j,k,m}=0$ and 0, otherwise. 
		\STATE Form a group per $\client_i$ using utility and costs $G_i=\{<\Utility_{i,m}, \cost_{i,m}, \video_{j,k,m}>\}$ where $m
\in\mathcal{M}_i.$
		\STATE Set partial solution: $G_{ps}=G_i$ and $\mathcal{G}=\bigcup_{\forall i\in\setRequests} G_i\setminus G_i.$
		\WHILE{$\mathcal{G} != \emptyset$} 
		\STATE Select a group  $G_{i}$ from $\mathcal{G}$ for merging with $G_{ps}.$
		\STATE 
		Initialize partial solution $G'_{ps}=\emptyset.$
		\FOR{$G_{ps,a}\in G_{ps}$   \label{alg:cph:merge-each-config-in-ps}} 
		\FOR{$G_{i,b}\in G_i$} \label{alg:cph:cartesian-1}
		\IF{$\video_{i,b}\in \video_{ps,a}$ or $\video_{i,b}==\video_{ps,a}$}  \label{alg:cph:video-check}
		\STATE $G_{ps,c} =<\Utility_{ps,a}+\Utility_{i,b}, \cost_{ps,a}, [\video_{ps,a}, \video_{i,b}]>$
		\ELSE
		\STATE $G_{ps,c} \label{alg:cph:diff-content} ={<\Utility_{ps,a}{+}\Utility_{i,b}, \cost_{ps,a}{+}\cost_{i,b}, [\video_{ps,a}, \video_{i,b}]>}$
		\ENDIF
		\STATE $G'_{ps}=G'_{ps}\bigcup G_{ps,c}$ 
		\ENDFOR
		\ENDFOR
		\STATE $\mathcal{G}=\mathcal{G}\setminus G_{i}$ and  $G_{ps}=G'_{ps}$
		\STATE Get Pareto-optimal points: $G_{ps}=\textrm{Pareto-min}(G_{ps})$. \label{alg:cph:pareto-min}
		\ENDWHILE
		\vspace{-10pt}
		\newtexthighlight{\IF{$G_{ps}=\emptyset$}
		\RETURN{$\widehat{\setRequests} =[\request_i]$.} \label{alg:cph:feasibility-check}
		\ELSE 
		\STATE Get the configuration with the maximum utility from $G_{ps}$ and retrieve the corresponding quality levels $\requestdelivered_i$.
		\RETURN{$\widehat{\setRequests}=[\requestdelivered_i]$.}
		\ENDIF}
	\end{algorithmic} 
\end{algorithm} 
Rather than considering all of the groups at once, CPH takes two groups and merges them simply by applying the Cartesian product.\footnote{\newtexthighlight{As shown in Fig.~\ref{fig:cph}, Cartesian product of $G_1$ and $G_2$ results in $G_1G_2$.}}~Afterward, in this set, CPH eliminates the configurations that are \textit{dominated} or that are infeasible due to a violation of resource constraints. \newtexthighlight{For instance, in Fig. \ref{fig:cph}, configuration (20, 850, [$r_2,r_2$]) dominates (7, 950, [$r_3,r_1$]) as its utility is higher while its cost is lower.}
At each step, CPH only keeps the  Pareto-optimal configurations. Such a step of reduction decreases the complexity of the problem significantly. CPH continues merging the set of Pareto-optimal points with one of the remaining sets. Optionally, to 
control its runtime and space complexity, at each intermediate step, CPH retains only ${L}$ configurations with the highest utility out of all Pareto-optimal points. 
After merging all sets, the configuration with the highest utility is selected. \newtexthighlight{In Fig. \ref{fig:cph}, the best decision is to assign quality level 2 to all clients which achieves a utility equal to 29 with resource consumption value of 1150.}  
While \cite{shojaei2013} discusses various approaches to optimize the merging step, e.g., to optimize the order of merging the sets, we do not consider any of the optional steps. 
Example in Fig. \ref{fig:cph} corresponds to our quality assignment problem when the clients request different contents and the AP's cache does not contain any of the candidate chunks.

In its original setting, CPH is not applicable to our formulated problem where the AP has a cache and there might be clients requesting the same content. Hence, we propose the following procedure to adapt CPH to out setting. The steps are are summarized in \textbf{Alg.~\ref{alg:cph}}.
First, we consider the cached content and assign their resource consumption to zero~(line \ref{alg:cph:cost}). 
\newtexthighlight{
As described before, CPH merges the groups by applying Cartesian product~(Line \ref{alg:cph:merge-each-config-in-ps} and Line~\ref{alg:cph:cartesian-1}) where utility and cost of the two members are added.} Nonetheless, in our problem, it is necessary to check if the two settings correspond to the same content~(Line~\ref{alg:cph:video-check}), so that we do not add the cost twice. This implies that the AP downloads each video item from the backhaul only once. In fact, this step breaks the main requirement of CPH that the utility and resource consumption of two configurations are additive. 
In case a content is requested by multiple clients, the MCKP abstraction might result in low utility. \newtexthighlight{If different contents are requested, we simply add the costs and utilities~(Line~\ref{alg:cph:diff-content}).}

Consider the example in Fig. \ref{fig:cph-modified} in which all clients request the same content. Notice that the utilities for the same quality level might differ from one client to another as the clients might have different buffer levels or channel link capacities. 
We modify CPH as follows: If a content is requested by multiple clients, we first merge these groups requesting the same content without Pareto elimination. Skipping the Pareto elimination step is crucial, as a  Pareto-dominated configuration in a single-user setting~(e.g., gray-shaded configurations in Fig. \ref{fig:cph-modified}) might yield the highest utility in a multi-user setting with a feasible cost. In Fig. \ref{fig:cph-modified}, the final solution achieving the maximum utility stems from one of the configurations that would have been eliminated if Pareto elimination had been applied. As Fig. \ref{fig:cph-modified} shows, CPH without any modifications would result in a lower utility: 27 as compared to 35. 
After all groups associated with this content are merged one by one, we reduce the set to Pareto-optimal points, since from now on we do not have the risk of eliminating potentially good solutions.
While this approach works for small settings, it increases the complexity significantly for some scenarios, e.g., in case several contents being requested by many clients. Therefore, at each intermediate step, only ${L}$ configurations with the highest utility can be kept for maintaining a higher scalability. However, tuning ${L}$ is not straightforward. Therefore, we prefer CPH with Pareto minimization step~(Line~\ref{alg:cph:pareto-min}). 
\newtexthighlight{
For cases where there is no feasible configuration, the AP does not overwrite the client requests~(Line \ref{alg:cph:feasibility-check}).} 

For a video with $\numRep$ quality levels, the worst-case complexity of CPH is $\mathcal{O}(\numClient\max(\numRep\log(\numRep),L^4))$~\cite{shojaei2013}. Note that the actual complexity is usually much lower, e.g., there are $2\toleratedQualityDiff+1$ quality levels in each configuration set rather than $\numRep$. 
\subsection{Airtime assignment for minimizing buffer stalls}
\label{sec:airtime-buffs}
After assigning the quality levels, i.e, $\widehat{\setRequests}=[\requestdelivered_i]$, we proceed with airtime assignment. 
To decrease the probability of buffer stalls, the AP aims at sustaining a minimum buffer level for its clients, e.g., $\bufferLevel_{\min}>0$. To this end, the AP calculates the required airtime to reach the aforementioned target level based on the current value $\bufferLevel_i$. However, if there is not sufficient content in the AP's downlink queue for some specific client, the allocated airtime is wasted. Hence, the AP shall consider the queue size for each client~($\queueSize_i$). Formally, the AP calculates the required airtime for each specific client $\client_{i}$ as follows: 
\begin{equation}
\airtime_i = \min(\queueSize_i, (\bufferLevel_{\min} - \bufferLevel_i)\bar{b}_i)/(\userRate_i T_{ap}),
\end{equation}
where $\bar{b}_i$ is the average bit rate of the chunks that are in the DL queue for client $\client_{i}$. If the required airtime is positive, the AP moves $\client_{i}$ to the list of clients that might experience buffer stalls, referred to as the set of \textit{risky} clients. In case the sum of all the required airtime by such clients exceeds 1, then the AP allocates each client some airtime which is proportional to its actual need normalized by the total required airtime of the risky clients.
The clients who have already sufficient media in their buffer~(e.g., two chunks are already in the buffer) are not served in this interval. \newtexthighlight{If the total required airtime for risky clients is less than 1, first each risky client receives its required airtime. Next, the AP divides the remaining airtime equally among the remaining clients not in the risky set.}

\section{Quality assignment for avoiding buffer-stalls~(\buffHeuristic)}
\label{sec:buff} 
As discussed in Section \ref{sec:MCKP}, in some cases~\newtexthighlight{(e.g., when multiple clients request the same content)}, CPH might have inferior performance compared to some heuristic that does not use the MCKP abstraction. \textit{To address this issue, we propose \buffHeuristic, whose objective is to assign a high video rate while avoiding buffer stalls.} As the cached chunks might be prioritized, \buffHeuristic~also uses a weighted sum as its objective: $\log(\mappingLeveltoBitrate_{j,\qualitydelivered_{i}}) (\weight_{c}\serveFromCache_i{+} (1{-}\serveFromCache_i))$. 
\begin{figure}[tb]
	\centering
	\includegraphics[width=0.5\textwidth]{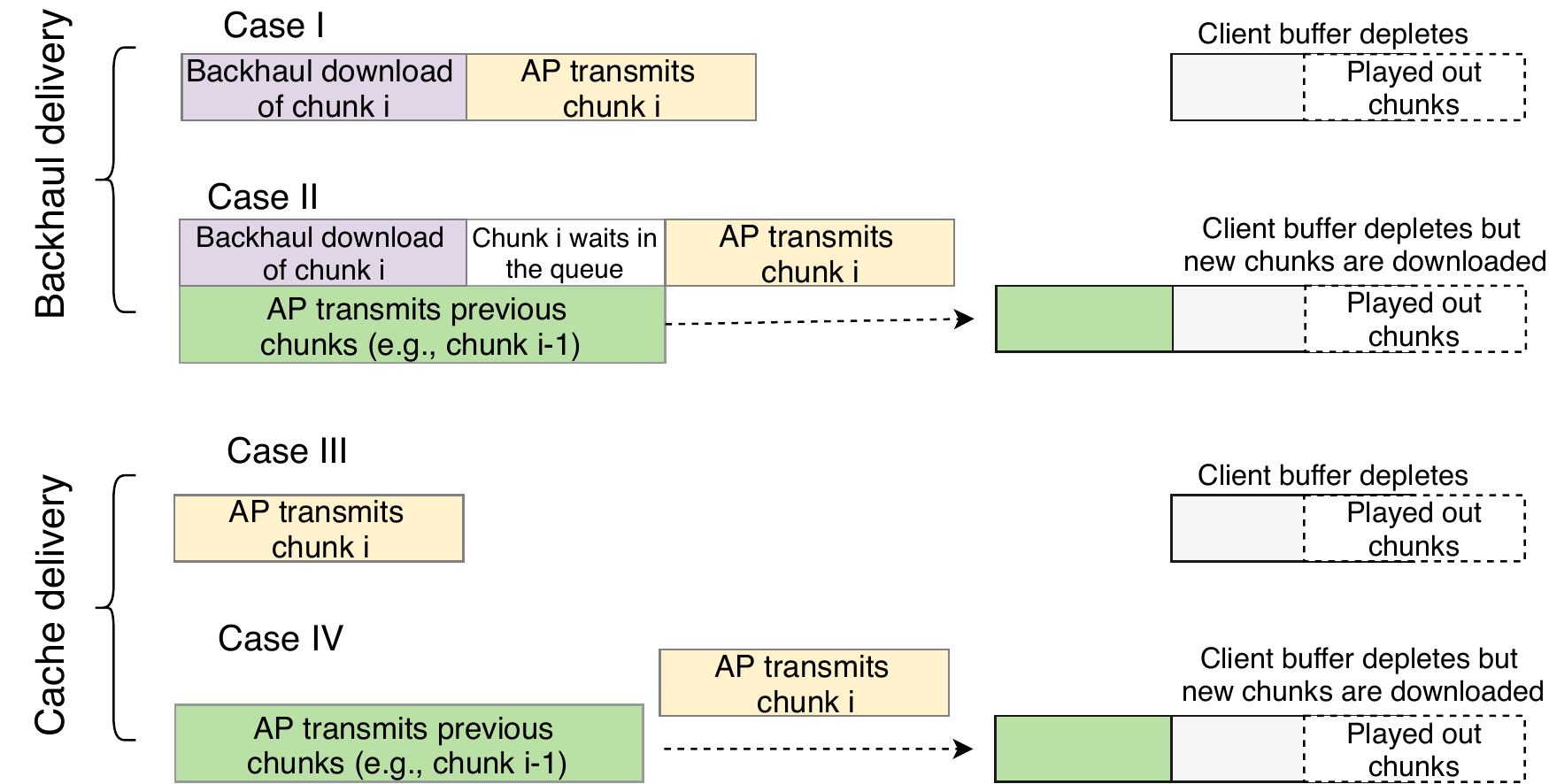}
	\caption{
		Illustration of the buffer dynamics while the current requested video is being downloaded from the backhaul. \textbf{Case I:} During the download, the AP waits as it has no bits to transmit to the client. Hence, the client's buffer will deplete. \textbf{Case II:} Client buffer both depletes with the rate of playout and increases with the rate of downloaded video from the AP's queue. Time needed to download the current c is shorter than the time needed to transmit all bits in the queue. As a result, the currently downloaded chunk experiences queuing latency in the AP's queue before it is delivered to the client. \textbf{Case III:} The content will be delivered directly from the cache. \textbf{Case IV:} Since the AP has other chunks to deliver to this client, the content fetched from the cache waits till the previous chunks are delivered to this client. 
		\label{fig:sixcases}}
\end{figure}
Let $\bufferInSecsEstimated_i$ denote the buffer level of client $\client_{i}$ when the currently requested chunk, e.g., chunk $i$, is downloaded to $\client_{i}$. $\bufferInSecsEstimated_i$ depends on three factors, namely, (i) 
current state of the buffer, (ii) from where the chunk is delivered~(i.e., backhaul or the cache), and 
(iii) the state of the current downlink queue of $\client_{i}$ at the AP.
Based on the combination of the aforementioned factors, a number of scenarios might occur. Below, we explain these cases which are illustrated in Fig. \ref{fig:sixcases}.
\begin{itemize} [leftmargin=0.1in]
	\item \textbf{Case I: Backhaul, empty client DL queue-} In this case, $\client_{i}$ requests some chunk $i$ that is not stored in the cache. Consequently, it has to be downloaded from the backhaul link that might also have other chunk requests waiting in the backhaul queue. Let $\Time_{b}$ denote the latency due to backhaul download, which depends on the quality of the chunk as well as the queue size at the backhaul FIFO queue. Moreover, the AP does not have any other chunks to deliver to $\client_{i}$, meaning that the DL queue of $\client_{i}$ is empty. The downloaded chunk is then transmitted in the DL according to the airtime allocated to $\client_{i}$. Let $\Time_{dl}$ indicate the required time to complete the transmission. Thus, in total, the delivery of the chunk to $\client_{i}$ takes $\Time_{b}+\Time_{dl}$ time units. 
	Meanwhile, the client buffer depletes, resulting in $\bufferInSecsEstimated_i{=}\bufferLevel_i{-}(\Time_{b}{+}\Time_{dl})$. As mentioned before, $\bufferInSecsEstimated_i$ can take negative values, reflecting the buffer stall period. 
	\item \textbf{Case II: Backhaul, non-empty client DL queue-} This case is similar to Case I, except for the fact that the client's DL queue is not empty. As a result, the AP transmits the existing chunks to the client while simultaneously downloading the chunk from the backhaul. Hence, on one hand, the buffer level decreases due to the play-out, and on the other hand, it increases by downloading the queued chunks. Therefore, we calculate the number of chunks that can be transmitted during $\Time_{b}$. Let $\queueSize_{i}$ and $\chunkduration_{\queueSize}$ indicate the queue size and the corresponding number of chunks, respectively. Then, the AP requires $\queueSize_i/(\userRate_i\airtime_i)$ seconds to transmit all bits in its DL queue. If the $\Time_{b}$ is shorter than this duration, the newly-downloaded chunk has to wait until all bits are delivered. Otherwise, the chunk does not experience any queuing delay in the downlink. 
	Consequently, the estimated buffer yields $\bufferInSecsEstimated_i=\bufferLevel_i-\max(\queueSize_i/(\userRate_i\airtime_i),\Time_{b})-\Time_{dl}+ \chunkduration_{\queueSize}$. 
	
	%
	\item \textbf{Case III: Cache, empty client DL queue-} In this case, it is clear that $\bufferInSecsEstimated_i=\bufferLevel_i-\Time_{dl}$. 
	\item \textbf{Case IV: Cache, non-empty client DL queue-} Here the AP first transmits the existing chunks in the DL queue. Subsequently, we calculate $\bufferInSecsEstimated_i$ as  $\bufferInSecsEstimated_i=\bufferLevel_i-\queueSize_i/(\userRate_i\airtime_i)-\Time_{dl}+ \chunkduration_{\queueSize}$.  
\end{itemize}
\begin{algorithm}[t]
	\textbf{\caption{
	\buffHeuristic
	}\label{alg:buff}}
	\begin{algorithmic} [1]
		\STATE \textbf{Input}: Requests to be assigned a quality~($\setRequests =[\request_i]$), client-AP link effective capacity, AP DL client queues~($\queueSize_i$), client buffer level~($\bufferLevel_i$)
		, available backhaul capacity~($\bottleneckCapacity$), cache status~($\cacheStatus=[\varcacheStatus_{j,k,m}]$)
		\STATE \textbf{Output}: $\widehat{\setRequests}$: video chunks to deliver for each $\client_i$.
		\STATE Initialize quality assignment list as $\requestdelivered=[]$.
		\STATE Find the candidate qualities for each client request using tolerated quality difference $\toleratedQualityDiff$. \label{alg:buff:tolerated_qualities}
		\STATE Calculate utility of each client and the quality pair $\Utility_{i,m}$. Add it to the set of utilities $\Utility$.
		\label{alg:buff:utility_calculation}
		\STATE Calculate the delivery costs $\cost_{i,m}$.
		\WHILE{$\setRequests !=\emptyset$ and $\bottleneckCapacity>0$}
		\STATE Get the best setting: $(i*,m*) = \arg\max_{i,m} \Utility$. \label{alg:buff:argmax}
		\STATE Assign quality $m*$ to the request of $\client_i*$: $\requestdelivered_{i*} = m*$ and the corresponding chunk is content $\video_{j*,k*,m*}$.
		\STATE Remove $\request_i*$ from unassigned requests, i.e., $\setRequests=\setRequests\setminus\request_i*$, and 
		all configurations of $\client_i$, i.e., $\Utility=\Utility\setminus\Utility_{i*,}$.
		\STATE Set the cost of all requests for the chunk $\video_{j*,k*,m*}$ to zero, i.e., $\cost_{l,m*}$=0, where $\request_l=\video_{j*,k*}$. \label{alg:buff:setcostzero}
		\STATE Decrease the available backhaul as $\bottleneckCapacity=\bottleneckCapacity-\cost_{m*}$. \label{alg:buff:decreasebackhaul}
		\STATE Remove infeasible settings with $\cost_{l,m}>\bottleneckCapacity$ from $\Utility$\label{alg:buff:bufferoverflow}.
		\STATE Append assigned quality $\requestdelivered_{i*}$  to $\widehat{\setRequests}$.
		\ENDWHILE
		\RETURN{$\widehat{\setRequests}$}
	\end{algorithmic} 
\end{algorithm} 

In the rest of this section, we describe our proposed algorithm, \textit{BUFF}. Moreover, we provide a brief summary of the procedure in \textbf{Alg. \ref{alg:buff}}.
The AP first finds all of the quality levels that are in the tolerated range of the client~(Line~\ref{alg:buff:tolerated_qualities}). Moreover, the AP analyzes the expected buffer level assuming identical airtime allocation. Since the goal is to avoid any buffer stall, the AP omits the quality levels that cannot fulfill this goal. 
However, if the quality level is the minimum level that can be assigned to the client, the AP keeps it as the only viable option. It then calculates the utility of each quality level~(Line~\ref{alg:buff:utility_calculation}). Afterwards, it greedily assigns the quality levels by picking the highest utility among all of the client-quality pairs~(Line~\ref{alg:buff:argmax}). After the AP assigns some client $\client_{i}$ a quality level, the cost of the remaining clients demanding the same chunk becomes zero since the AP does not download a content multiple times~(Line~\ref{alg:buff:setcostzero}). \newtexthighlight{The AP decreases the available backhaul capacity considering the bitrate assigned in this step~(Line~\ref{alg:buff:decreasebackhaul})} and removes the quality levels whose bitrate exceed the available remaining backhaul capacity~(Line~\ref{alg:buff:bufferoverflow}). 
\buffHeuristic~terminates either when at least one of the following conditions holds: (i) All clients are assigned a quality level; (ii) The backhaul capacity is exhausted. 
Finally, \buffHeuristic~uses the same airtime assignment approach introduced in Sec.\ref{sec:airtime-buffs}. 

The computational complexity of \buffHeuristic~is calculated as follows.  
For each client, \buffHeuristic~calculates the utility for all tolerated quality levels resulting in $\min(2\toleratedQualityDiff+1, \numRep)\numClient$ operations. Afterward, it finds the maximum utility at each iteration. The iterations continue until either all clients are assigned a quality level or the backhaul is exhausted. Assuming the first case occurs earlier, then the complexity yields $\mathcal{O}(\min(2\toleratedQualityDiff+1, \numRep)\numClient^2)$.
\section{Performance Evaluation}
\label{sec:eval}
To evaluate our proposals, we conduct intensive simulations using our system-level WiFi simulator developed in Python.\footnote{\newtexthighlight{Source code of the simulator can be provided on request.}}
\subsection{Evaluation setting}

The WiFi BSS operates on a channel of 40 MHz and we model the AP-client link as a Keenan-Motley channel~\cite{tornevik1993propagation}. At the client-side, the adaptation algorithm is the rate-based adaptive~(RBA) algorithm~\cite{mangla2016video}. Briefly, the client selects the highest bitrate smaller than the estimated rate. As rate estimation approach, we use harmonic-rate estimation algorithm~\cite{sun2016cs2p} which computes the harmonic mean of the previously downloaded five chunks.
For the initial chunks, there is no data to estimate the rate; therefore, the client picks the lowest bitrate until it fills out its buffer. In essence, most of the DASH clients follow this approach to ensure low startup latency. A client can generate back to back requests for the chunks to fill its buffer quicker in the initial phase, i.e., before watching the first chunk of the video.  After the client fills the buffer and playout starts, the client can have at most three requests on the fly. Moreover, when the buffer is full, the client does not generate any new requests until the buffer has some  space for the new chunks.

\begin{figure*}[tb]
	\centering
	\subfloat[Video quality in Kbps.]{\includegraphics[width=0.33\linewidth]{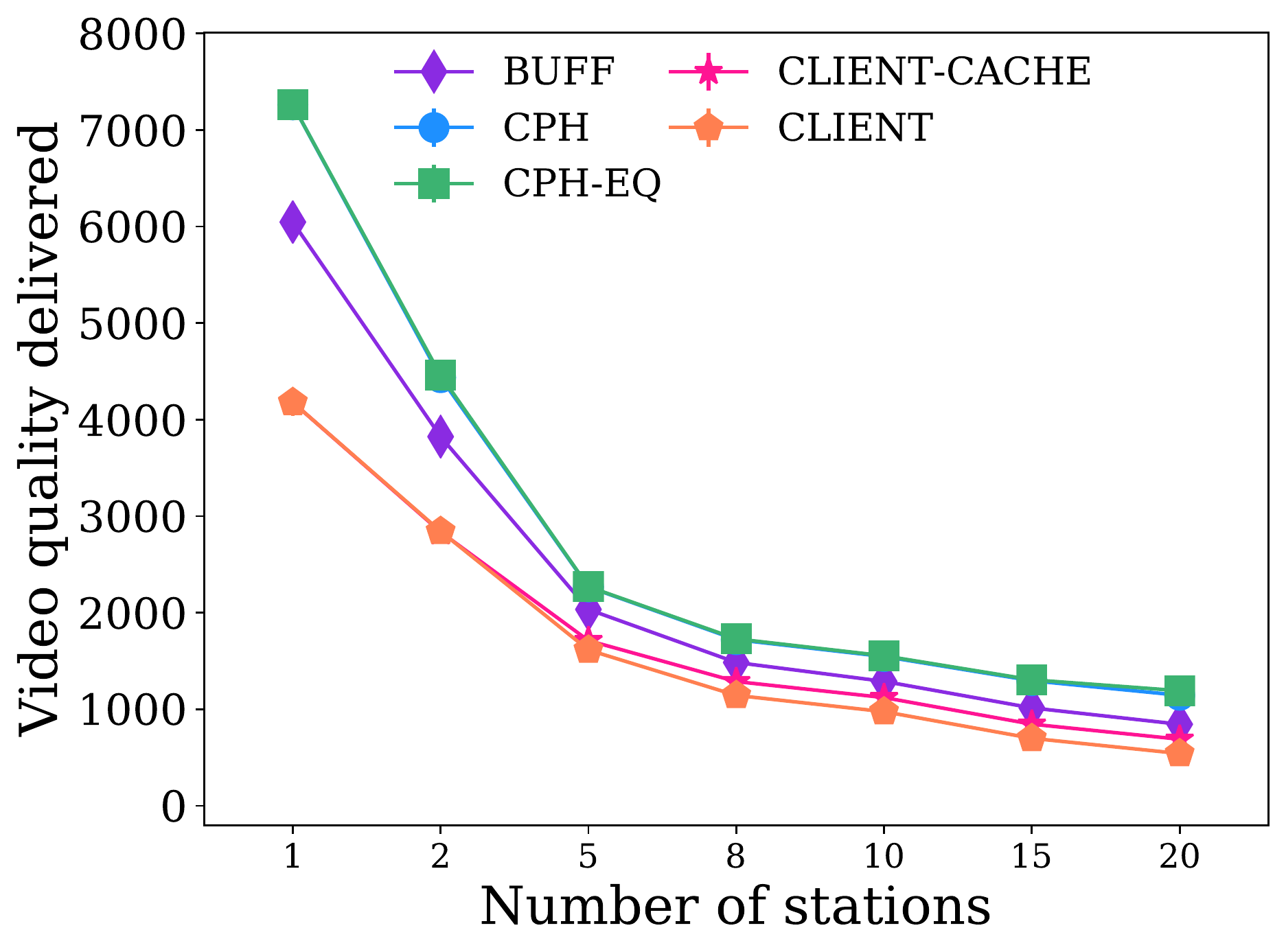}\label{fig:nclients-videoquality}}
	\subfloat[Cache bit hit ratio.]{\includegraphics[width=0.33\linewidth]{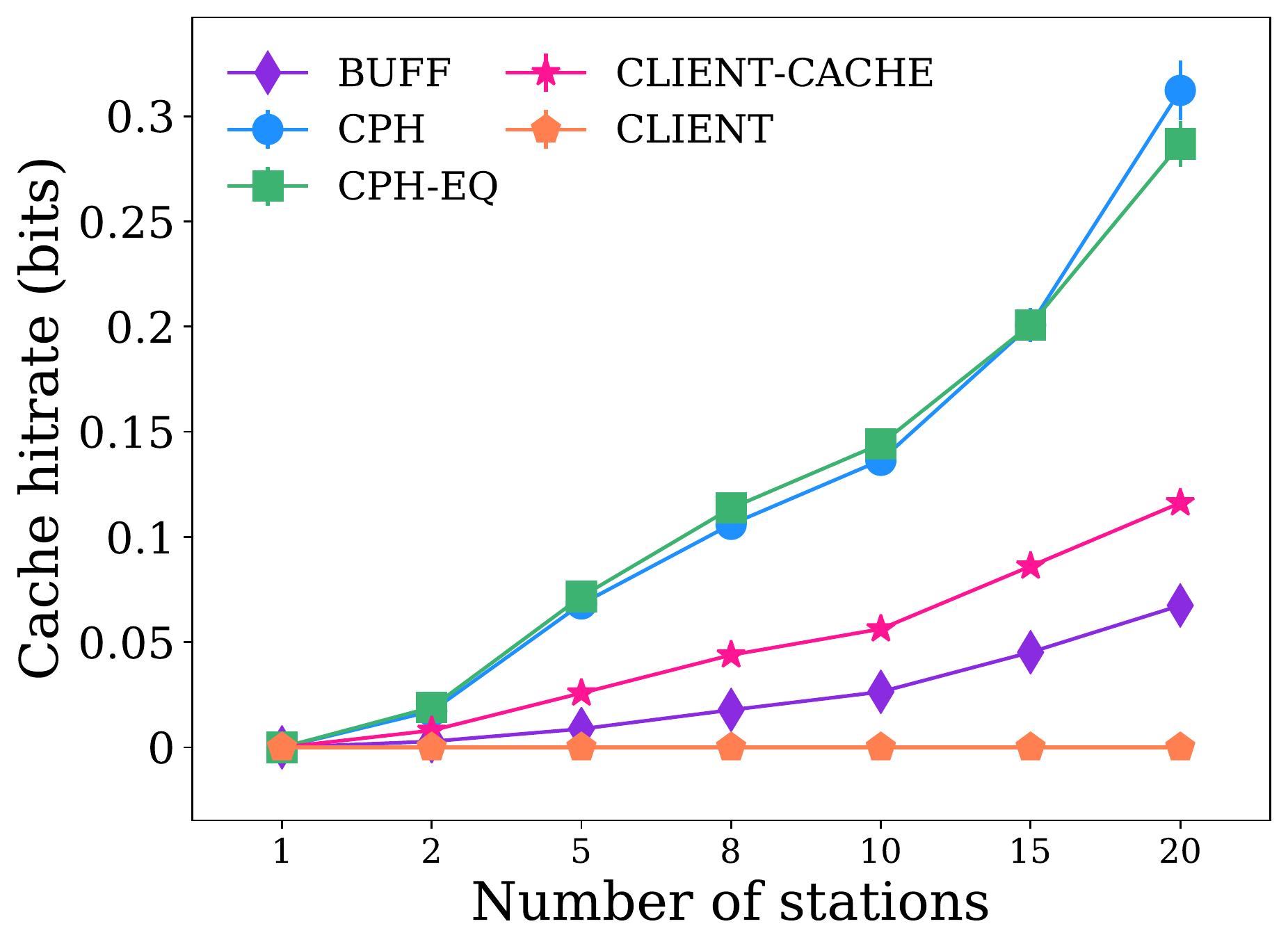}\label{fig:nclients-cachehit}}
	\subfloat[Stalling ratio.]{\includegraphics[width=0.33\linewidth]{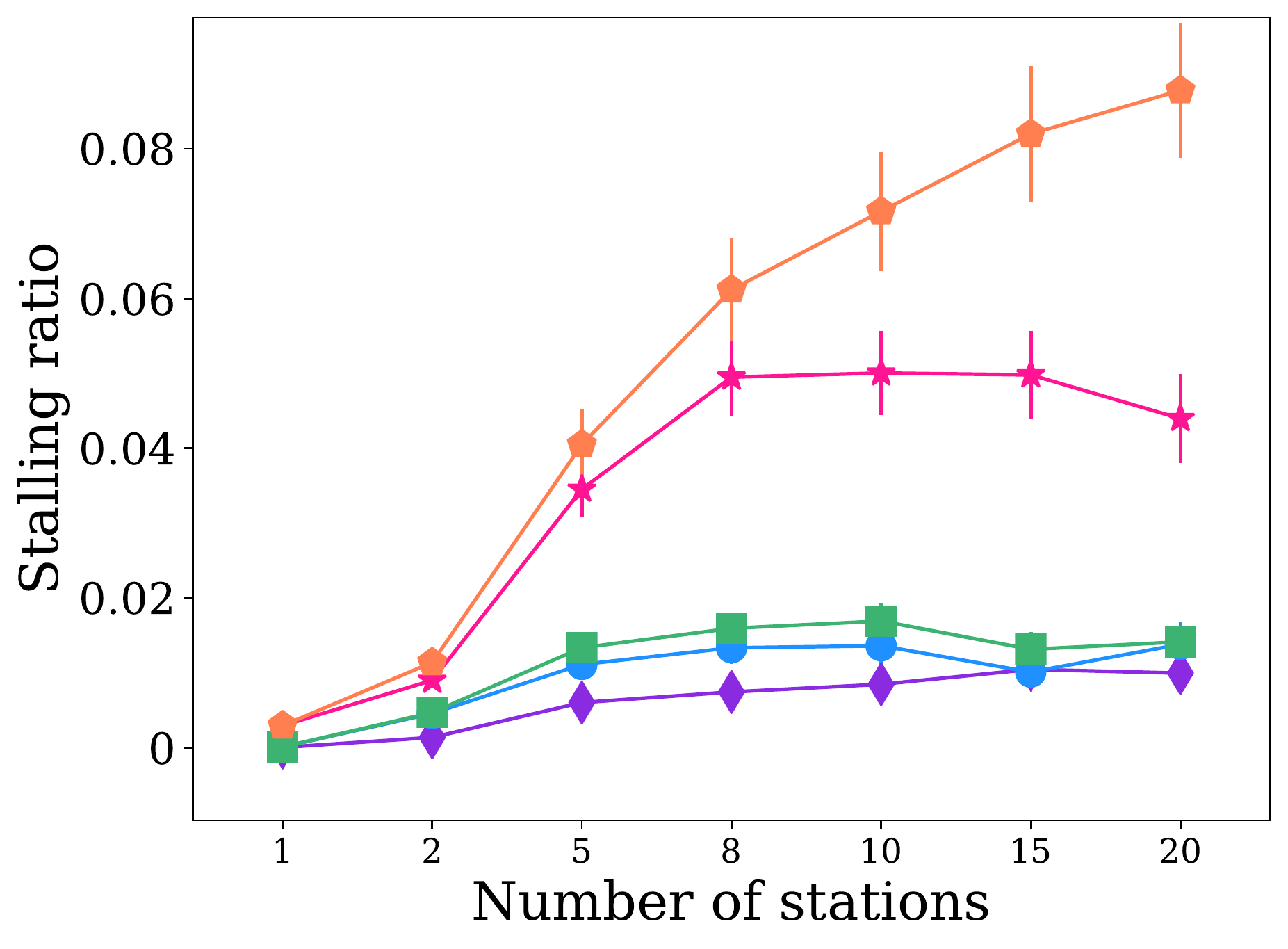}\label{fig:nclients-Stallingratio}} 
	\caption{Impact of the number of DASH clients on the video quality, the cache bit hit ratio, and the stalling ratio. 
 \label{fig:numClients}}
	\vspace{-10pt}
\end{figure*}

As our video content catalogue, we used both a real video trace~\cite{quinlan2016datasets} and synthetic video set.
The data set in \cite{quinlan2016datasets} has 23 video clips, each with either 16 mins or 10 mins length. 
Each video is encoded using H.264 encoder, considering the chunk durations of [2, 4, 6, 8, 10] seconds. We follow the recommendation of some earlier work\footnote{Please see more at  \url{https://bitmovin.com/mpeg-dash-hls-chunk-length/}.} to use chunk sizes of 2-4 seconds that finely addresses the trade-off between overheads and throughput efficiency. The average video bitrates are as follows: [232, 374, 560, 751, 1059, 1773, 2363, 3022, 3822, 4273] Kbps. Each video file records the actual chunk size information. 
We also generate synthetic traces with higher bitrates, e.g., from 100 Kbps to 15 Mbps, and with total 19 quality levels. 
Since the trends are similar, we report
results from the synthetic trace. 
We assign each client a video randomly from the video content set considering Zipf content popularity model with exponent 1.2.
We assume the existence of a high-capacity cache to keep the impact of the cache admission and replacement policy minimal.
Unless otherwise stated, we use the following parameters: $\toleratedQualityDiff=2$, $\bufferLevel_{\max}=15$ seconds and $\bufferLevel_{min}= 4$ seconds, $\weight_{c}= 1.3$, $\bottleneckCapacity= 20$ Mbps, $\numClient=10$, $\numVideo=10$, $T_{ap}=0.5$ seconds, and $\text{RAI}=T_{ap}$. Clients are uniformly distributed in the coverage area of the AP, which we model as a circle with its radius being 70 meters.

We evaluate the following schemes:
\begin{itemize}
	\item \textbf{CPH-EQ:} CPH with equal airtime allocation, 
	\item \textbf{CPH}: This approach is CPH with an airtime allocation that takes the buffer occupancy of each client into account, 
	\item \textbf{BUFF}: BUFF with airtime allocation identical to CPH,
	\item \textbf{CLIENT:} The AP acts as a repeater without checking if the requested content is already cached or not. Moreover, it allocates the airtime equally among its clients, and 
	\item \textbf{CLIENT-CACHE:} This scheme is similar to CLIENT, with the difference that here the AP can deliver the requested content from the cache if stored in the cache. 
\end{itemize}
Among the aforementioned schemes, CLIENT is the baseline scheme, as it corresponds to the usual operation of client-driven DASH. When CPH variants and BUFF cannot find a feasible solution, the AP does not change the requests and delivers the requested qualities. 
We run each scenario for 200 times and report the average of the statistics along with 95\% confidence intervals.

\subsection{Impact of the number of clients}

\begin{figure*}[tb]
	\centering
	\subfloat[Stalling ratio.]{\includegraphics[width=0.33\linewidth]{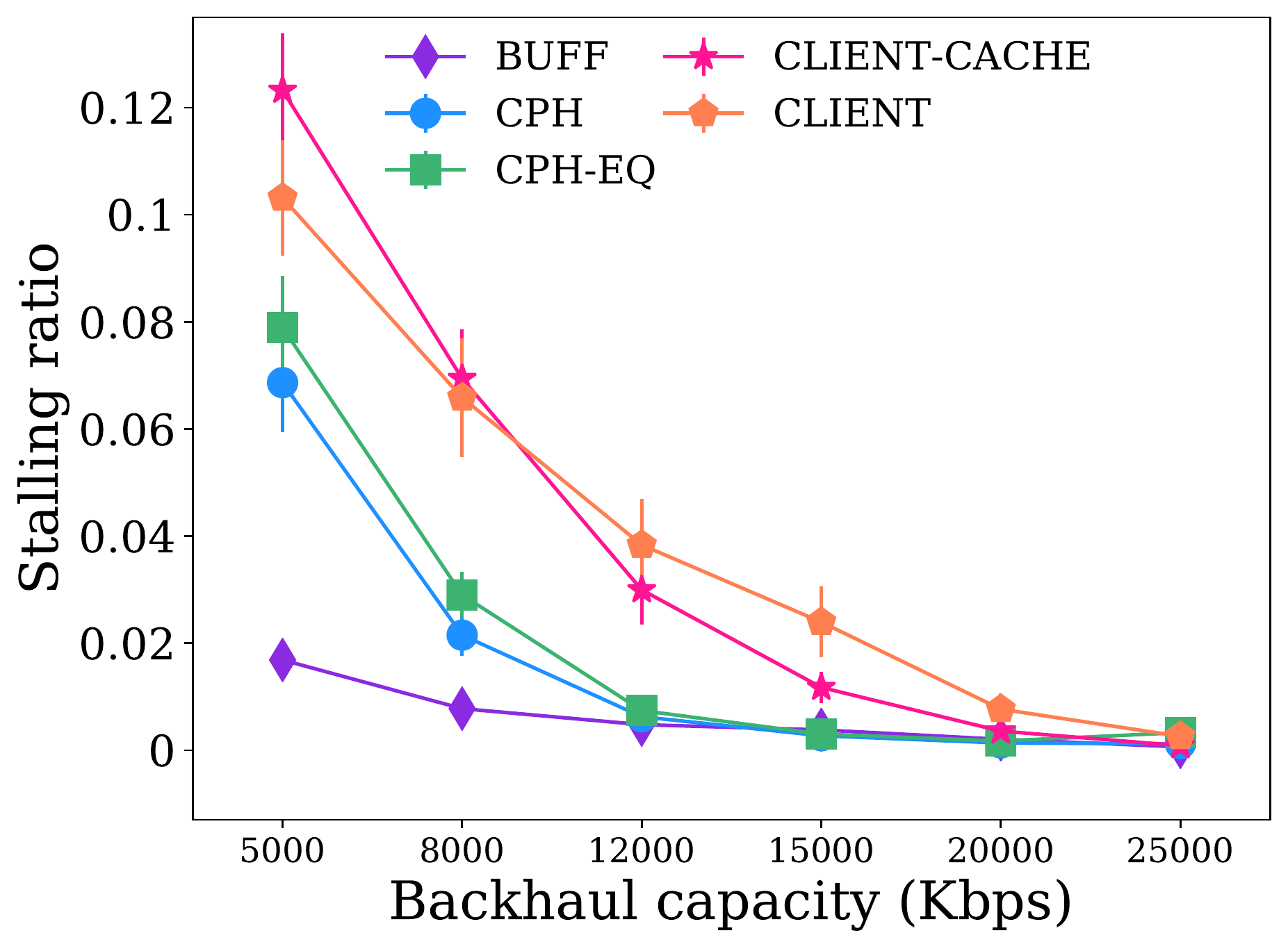}\label{fig:bh-Stallingratio}} 
	\subfloat[Video quality in Kbps.]{\includegraphics[width=0.33\linewidth]{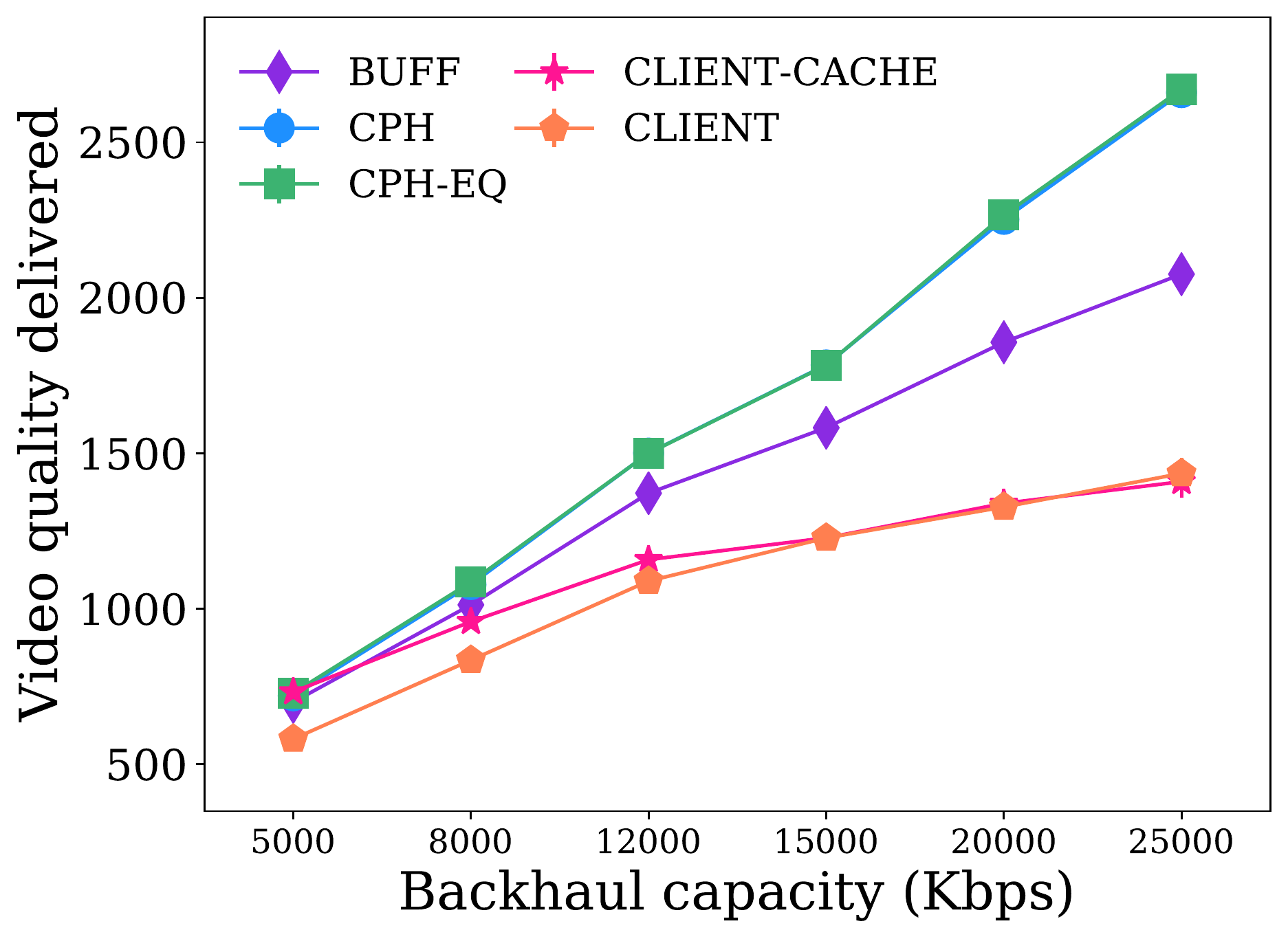}\label{fig:bh-videoquality}}
	\subfloat[Cache bit hit ratio.]{\includegraphics[width=0.33\linewidth]{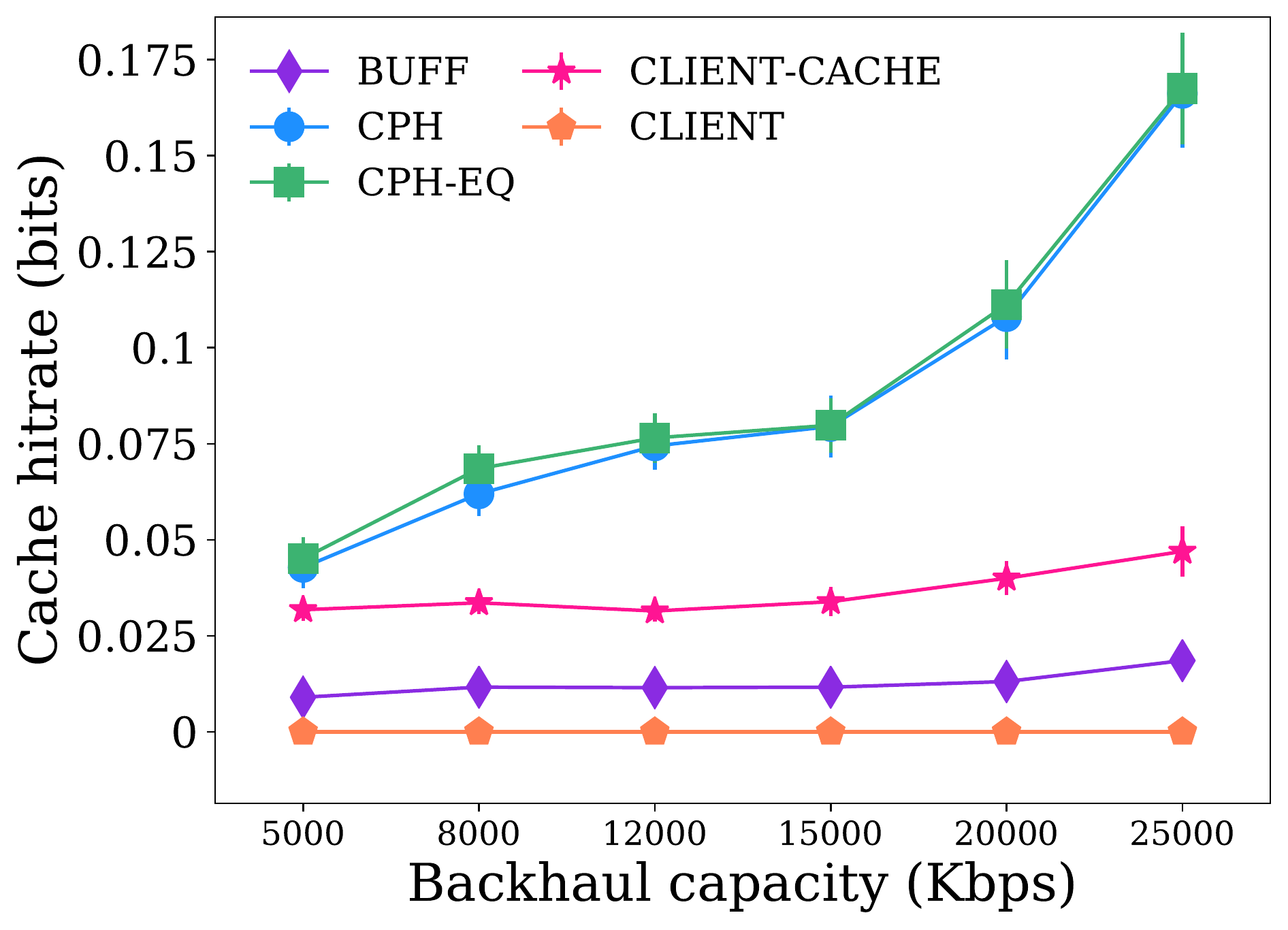}\label{fig:bh-cachehit}}  \\
	\subfloat[Initial latency in seconds.]{\includegraphics[width=0.33\linewidth]{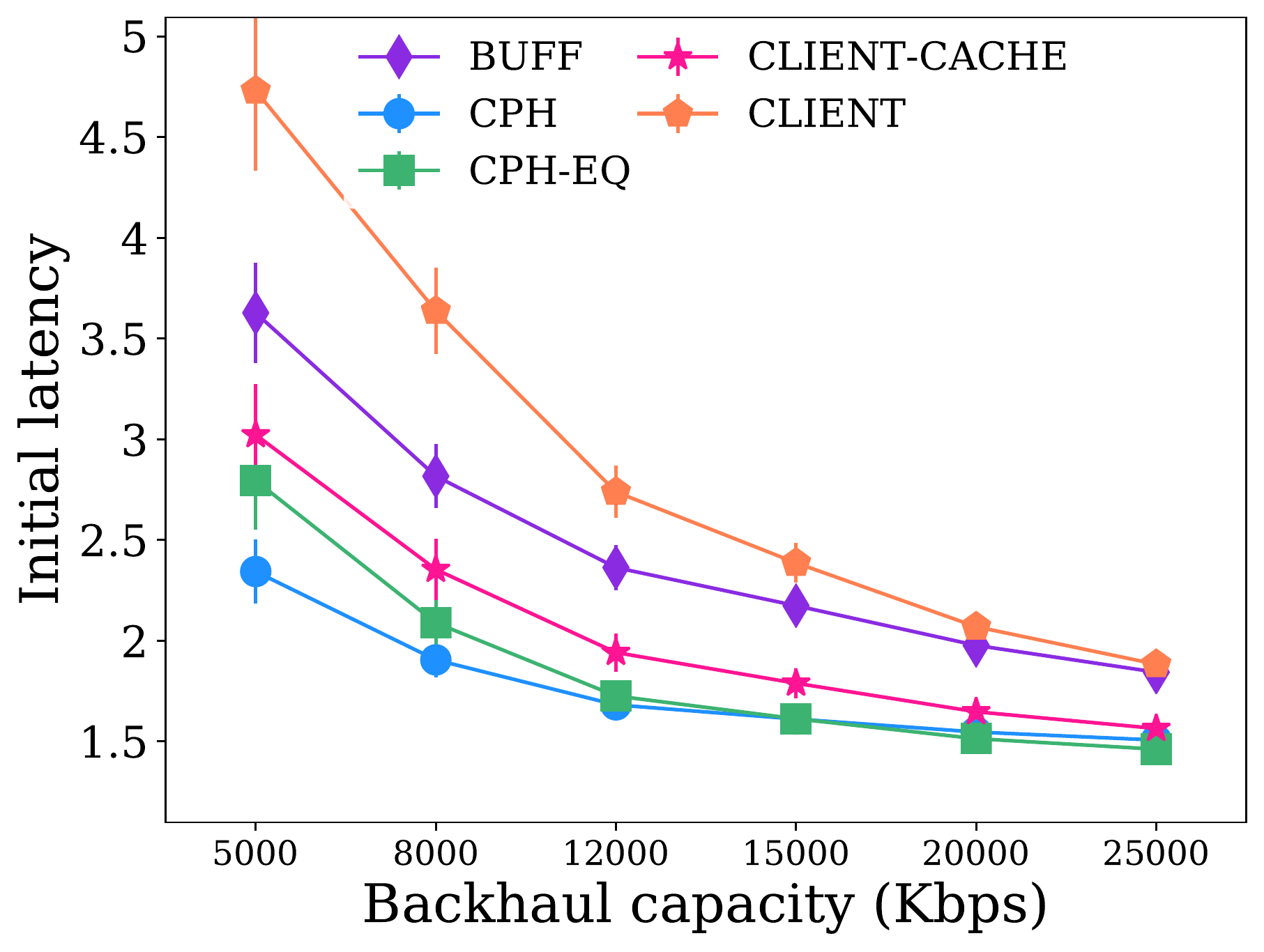}\label{fig:bh-Initiallatency}}
	\subfloat[Backhaul utilization.]{\includegraphics[width=0.33\linewidth]{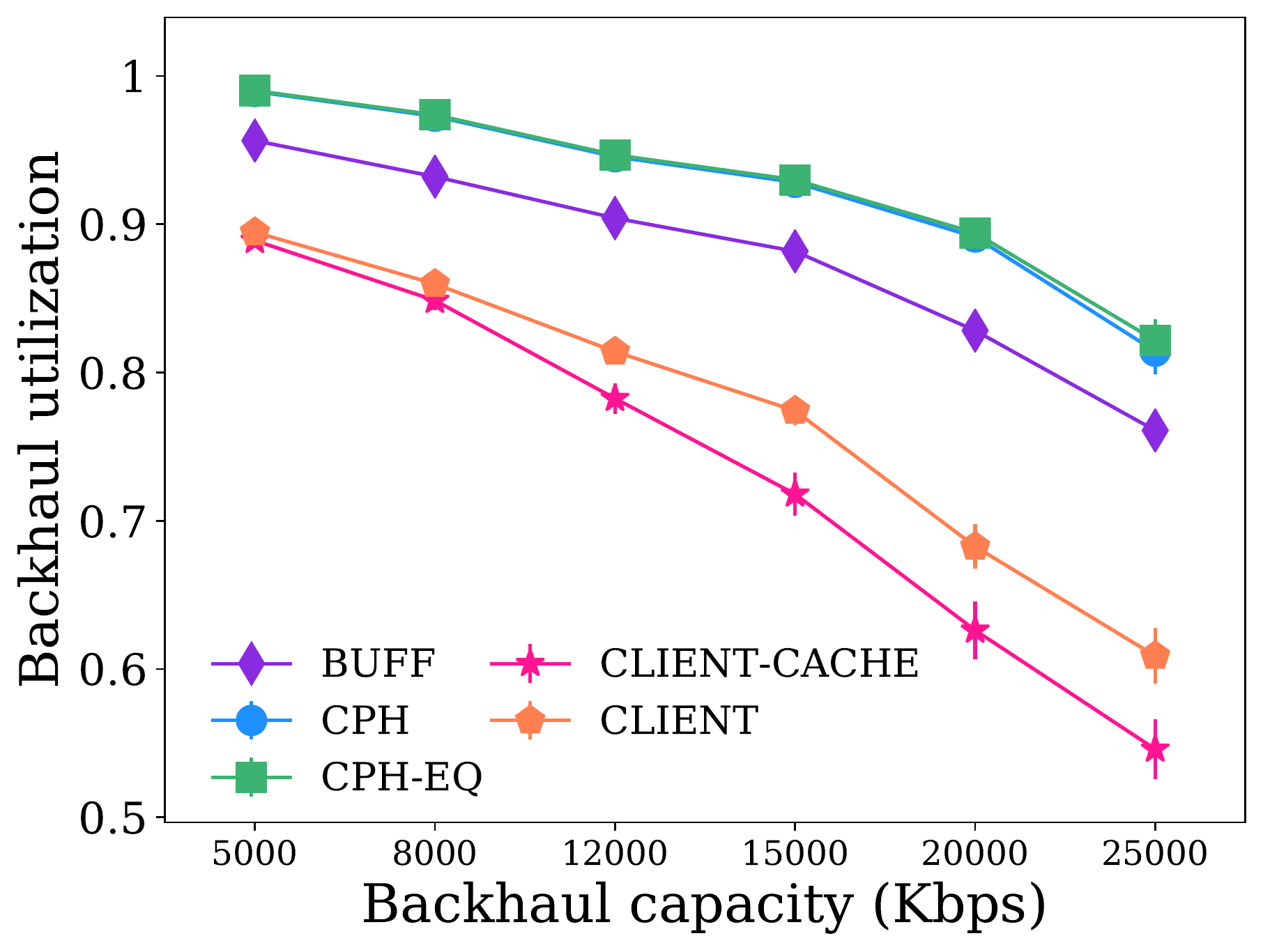}\label{fig:bh-Backhaulutilization}} 
	\subfloat[No valid configurations.]{\includegraphics[width=0.33\linewidth]{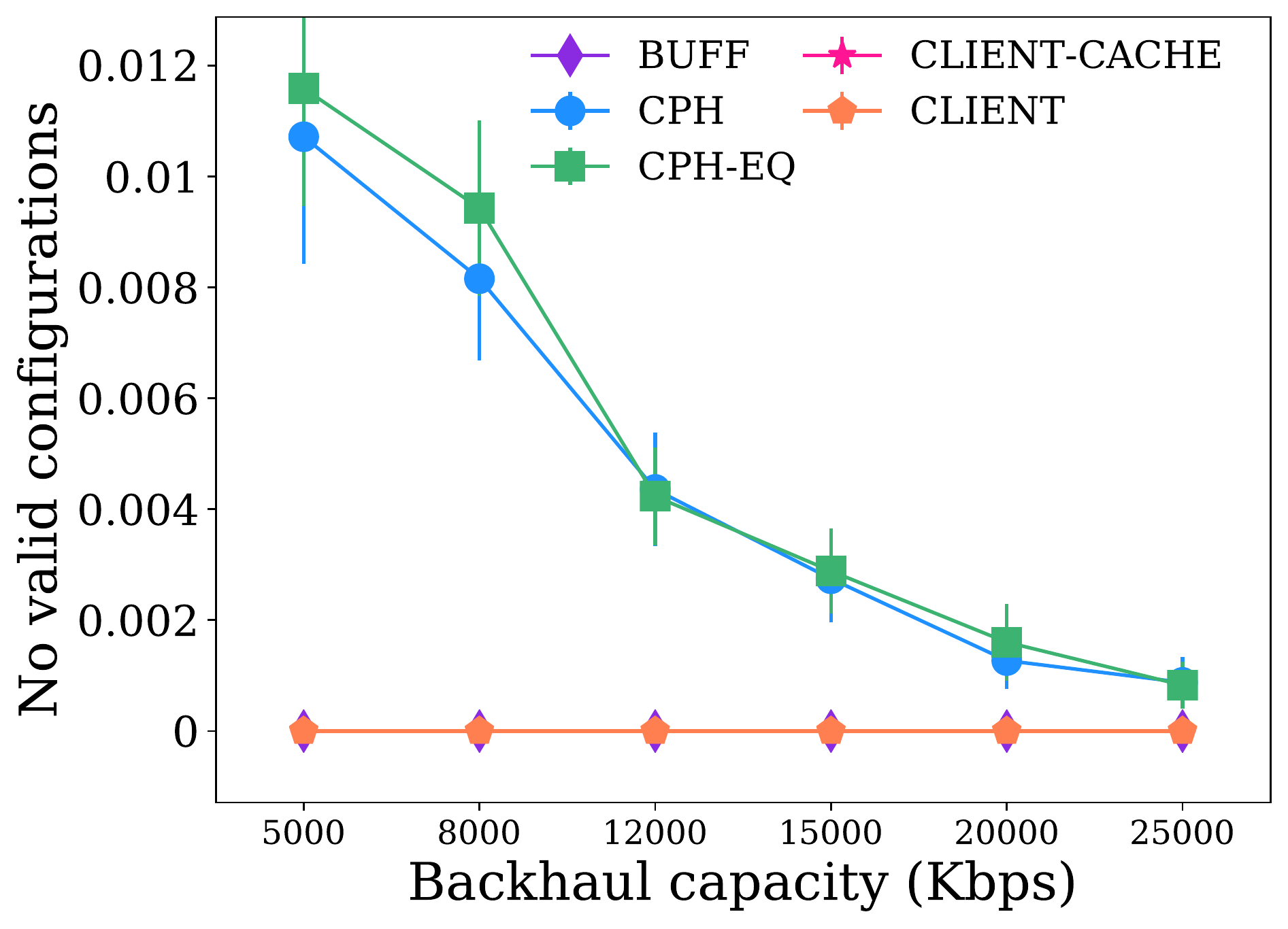}\label{fig:bh-novalid}} 
	\caption{Impact of increasing backhaul capacity under $\numClient=10$, synthetic video data, 10 videos. \label{fig:backhaul}
	}
	\vspace{-10pt}
\end{figure*}
Fig. \ref{fig:numClients} shows the performance of each scheme as a function of the number of DASH clients ($\numClient$). 
In Fig.~\ref{fig:nclients-videoquality}, we observe that all schemes maintain a lower video bitrate with larger $\numClient$ while CPH variants sustain the highest video bitrates without resulting in many stalls~(Fig. \ref{fig:nclients-Stallingratio}). 
For example, for a single user, CPH provides 3 Mbps higher bitrate corresponding to 74\% improvement over CLIENT, whereas
the improvement is 112\% when $\numClient=20$ enabled by 0.3 Mbps higher bitrate. Similarly, BUFF provides 45\% and 56\% improvement for $\numClient=1$ and $\numClient=20$, respectively. 
With increasing $\numClient$, the stall ratio increases for all schemes. However, the growth shows a higher rate for the variant of CLIENT such that the video session might become very unpleasant. As a comprehensive example, consider the following scenario. With only a few clients, streaming is smooth without interruptions for all schemes, as shown by Fig. \ref{fig:nclients-Stallingratio}; nevertheless, for $\numClient=5$, CLIENT results in stalls more frequently, specifically around 3-4\% of the session. With larger values of $\numClient$, stalls might occur even more often, as frequent as 6-8\% of the session, whereas it remains around 1\% for our proposals.

With respect to the video bitrate performance, the schemes can be sorted as follows: CPH or CPH-EQ, BUFF, CLIENT-CACHE, and CLIENT. There is almost no quality difference between CPH and CPH-EQ while we observed a slightly higher cache bit hit ratio achieved by CPH-EQ for large $\numClient$ and for our synthetic video trace~(omitted due to the space concerns). 
For cases where CPH-EQ has a higher cache hit ratio than that of CPH, it could be due to the similar link capacity observations of the clients, which result in requesting the same video qualities. 
Moreover, we believe that CPH-EQ can maintain the same performance as that of CPH because of the high capacity of the AP-client links. As opposed to the backhaul link whose utilization is around 90\%, the client-AP capacity is sufficient to serve all of the clients without resulting in long queues at the AP. In this scenario, the AP-client perceived link capacity is around 5-38 Mbps per client. Note that the perceived capacity depends on the activities of all of the clients since there might be some time intervals where the clients are in the OFF-state in the well known ON-OFF cycle of the DASH~\cite{bentaleb2018survey}.
Under high AP-client link capacity, the airtime allocation affects the performance only marginally, as confirmed by no performance difference between CPH and CPH-EQ for low $\numClient$. However, for larger $\numClient$, as Fig. \ref{fig:nclients-Stallingratio} shows, CPH maintains slightly lower stalling ratio than CPH-EQ.
Moreover, although BUFF achieves the lowest buffer stalls, it comes at the expense of lower video quality compared to CPH-variants and lower cache hits compared to CLIENT-CACHE.

Despite having network control, our schemes may suffer from the same problems as the client-driven approaches. The problems arise since the bitrate associated with a quality level is only an average value, which might differ from the actual chunk size. For example, the actual chunk size might be much larger than the one calculated using the denoted bitrate which then requires a longer time to download from the backhaul and to transmit to the client. Note that CLIENT-CACHE does not assign a quality level, and only delivers the requested content from the cache upon availability. 
Despite enabling cache delivery, this capability without quality assignment does not suffice to improve client performance as we observe high stalling ratio in 
Fig.~\ref{fig:nclients-Stallingratio}. With increasing $\numClient$, CLIENT-CACHE starts to find content in the cache and therefore we observe a slight decrease in the stalling ratio in Fig.~\ref{fig:nclients-Stallingratio}. 

\begin{figure*}[tb]
	\centering
	\subfloat[Video quality in Kbps.]{\includegraphics[width=0.33\linewidth]{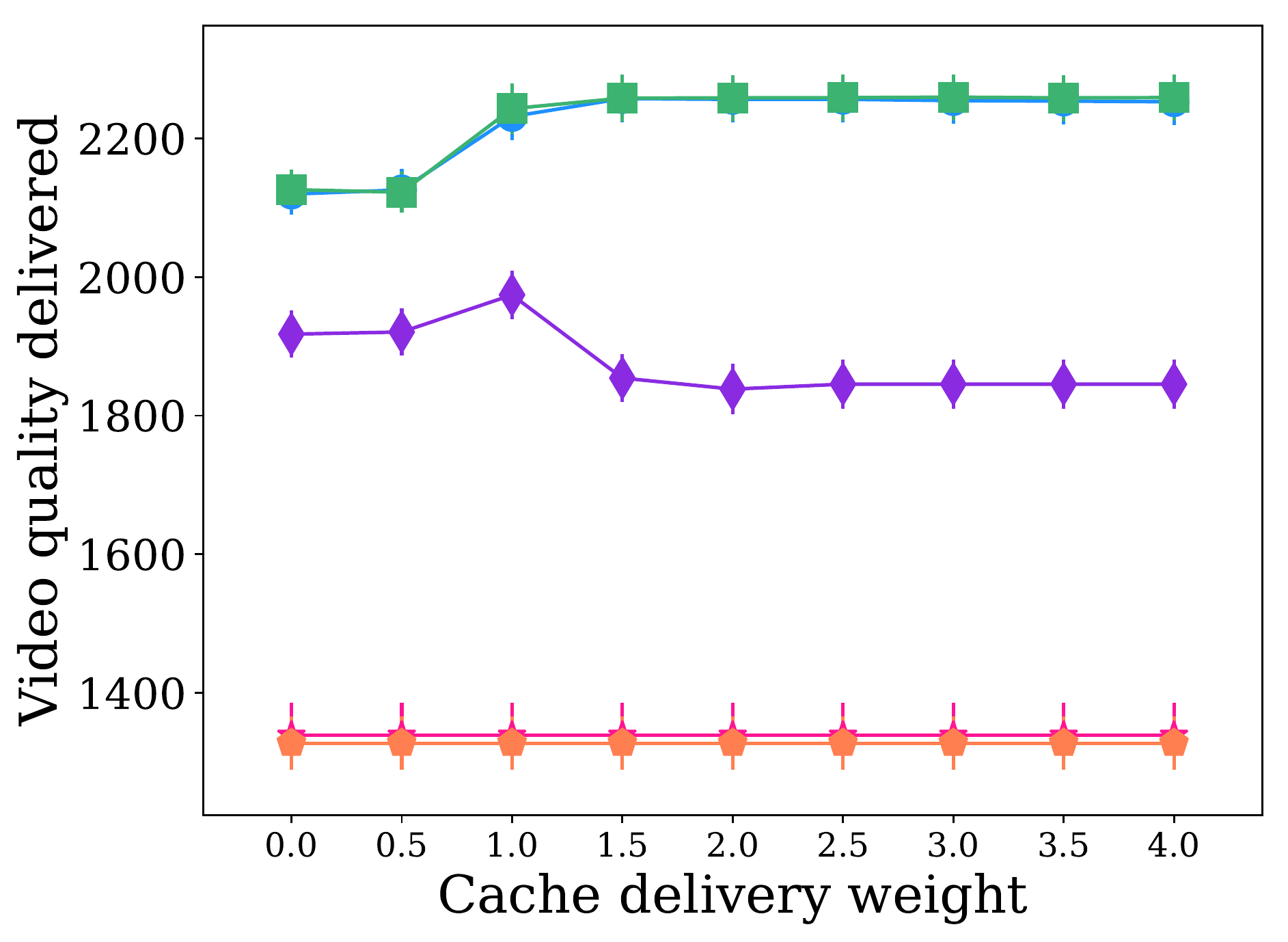}\label{fig:mu-videoquality}}
	\subfloat[Cache bit hit ratio.]{\includegraphics[width=0.33\linewidth]{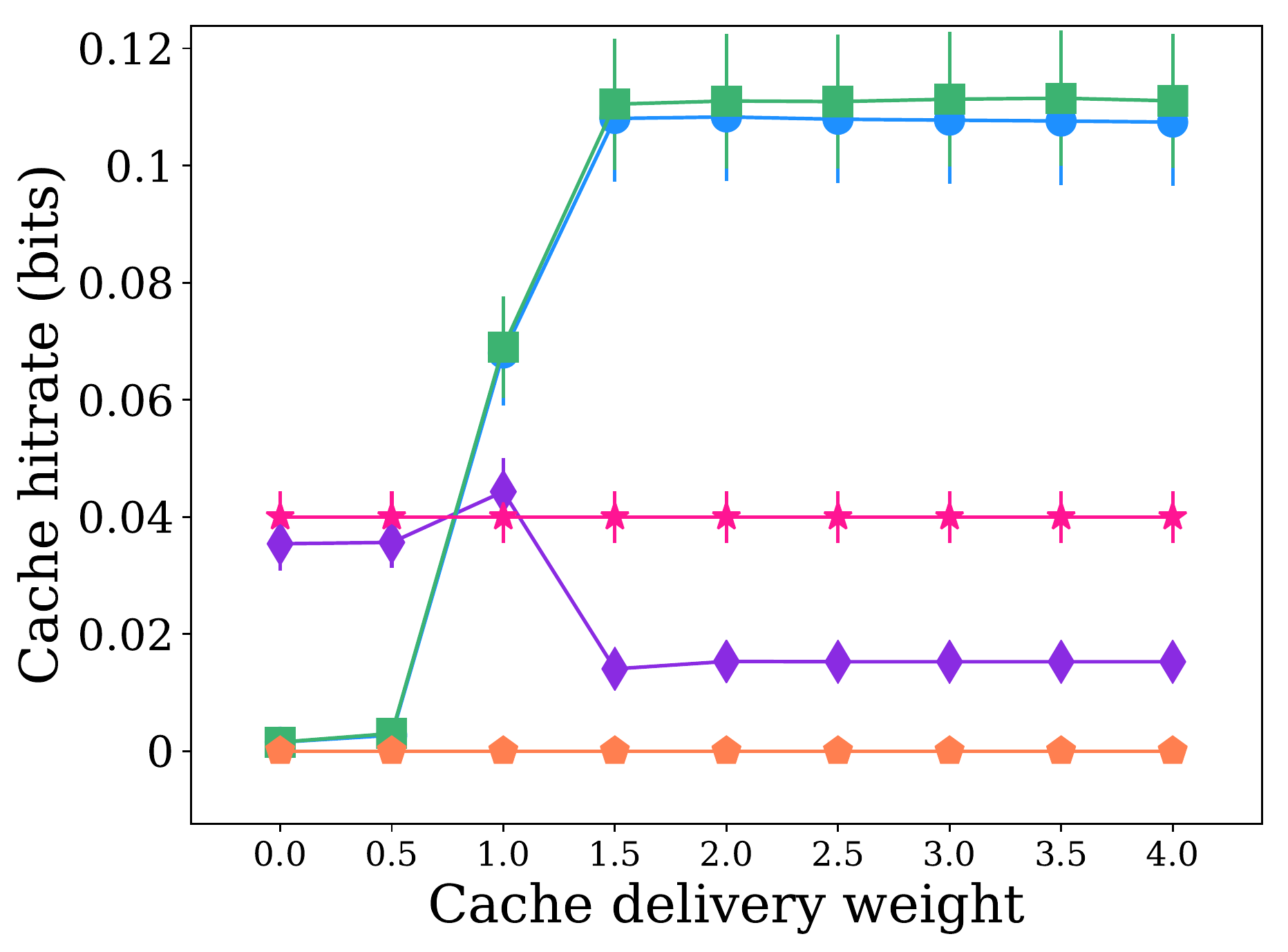}\label{fig:mu-cachehit}}
	\subfloat[Backhaul utilization.]{\includegraphics[width=0.33\linewidth]{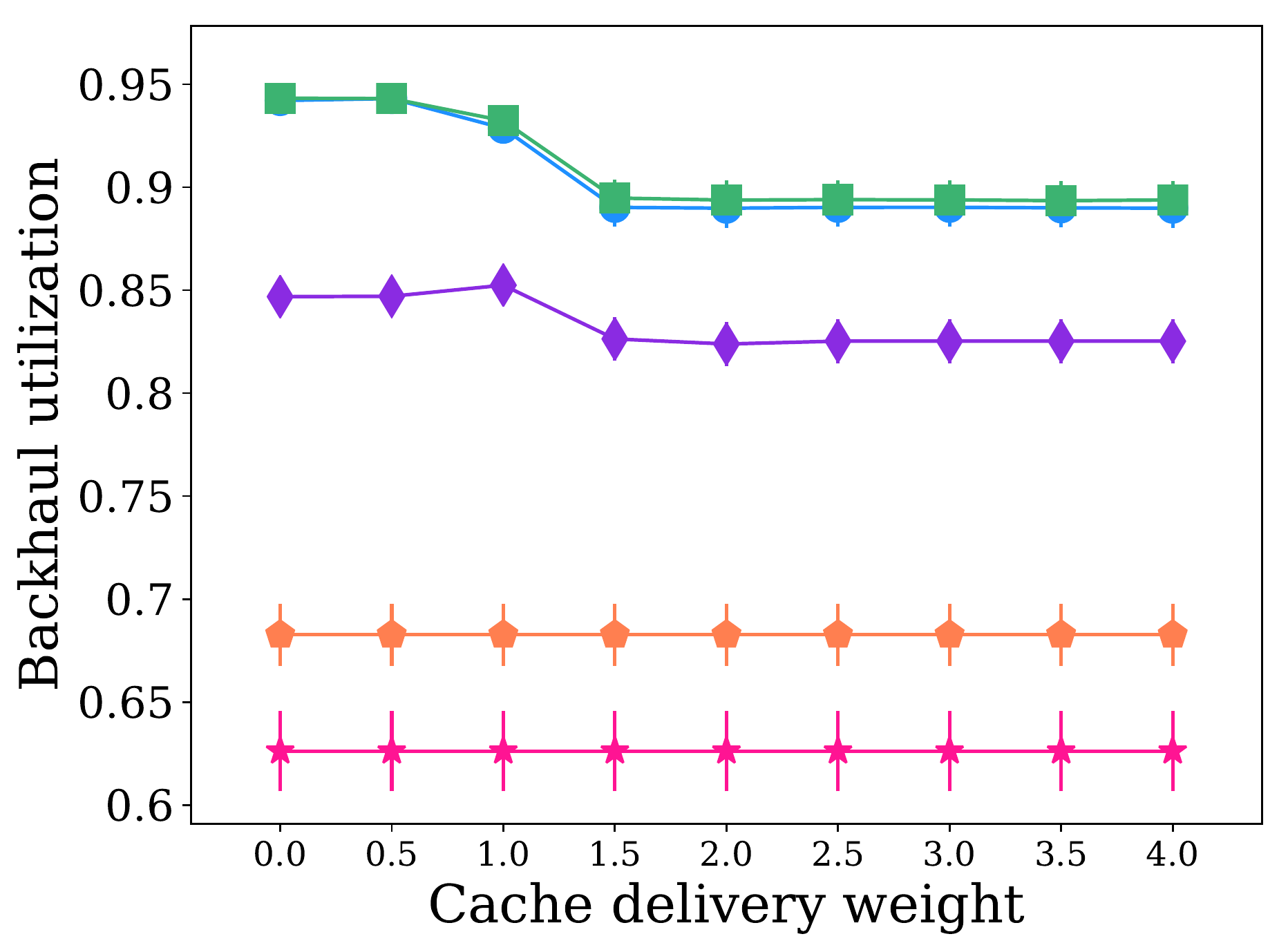}\label{fig:mu-Backhaulutilization}} 
	\caption{Impact of cache weight $\weight_{c}$, synthetic video files.
		\label{fig:cachemu}}
	\vspace{-10pt}
\end{figure*}

In summary, for a large number of DASH clients, CPH achieves a significantly higher cache hit ratio while simultaneously providing the highest video bitrate and keeping the buffer stalls very close to that of the BUFF. 

\subsection{Impact of the backhaul capacity}
In this scneario, to analyze the impact of the backhaul capacity $\bottleneckCapacity$, we fix the number of clients to 10. As Fig. \ref{fig:backhaul} shows, for all backhaul capacity settings, CPH and CPH-EQ outperform the variants of CLIENT and BUFF in most of the performance metrics.
For example, when backhaul capacity is abundant~($\bottleneckCapacity>20$\,Mbps), the stalling ratio is zero for all schemes in Fig. \ref{fig:bh-Stallingratio}. Nonetheless, the video bitrates are lower for CLIENT variants and BUFF~(Fig. \ref{fig:bh-videoquality}). 

Comparing BUFF and CPH, we can argue that BUFF is a better choice when $\bottleneckCapacity=5$\,Mbps as it sustains a lower stalling ratio compared to CPH and CPH-EQ.
In all other cases, CPH and CPH-EQ should be the choice not only for leveraging the cached content~(Fig. \ref{fig:bh-cachehit}) but also for a shorter initial latency~(Fig. \ref{fig:bh-Initiallatency}).
We also report backhaul utilization ratio in Fig. \ref{fig:bh-Backhaulutilization} and the fraction of time when CPH could not find a valid configuration in Fig. \ref{fig:bh-novalid}.
As Fig.~\ref{fig:bh-Backhaulutilization} shows, the CPH variants and BUFF could utilize the backhaul capacity better while CLIENT variants leave the capacity underutilized presumably due to an inaccurate estimation of the link capacity.
\subsection{Impact of the cache delivery weight}
Fig.~\ref{fig:cachemu} shows the impact of increasing cache delivery weight~$\weight_{c}$ when $\bottleneckCapacity$=20 Mbps.
CPH variants benefit from increasing $\weight_{c}$ from 1 to 1.5 slightly in terms of higher video quality and significantly regarding cache hits as shown in Fig. \ref{fig:mu-videoquality} and Fig. \ref{fig:mu-cachehit}.
The performance becomes stable after $\weight_{c}>1.5$. 
Similarly, Fig.\ref{fig:mu-Backhaulutilization} shows that increasing $\weight_{c}$ decreases the backhaul utilization first but stabilizes afterward. 
In this setting with 20 Mbps backhaul capacity, almost all schemes sustain a smooth streaming session without a buffer stall. But, under a bottleneck capacity of $\bottleneckCapacity$=8 Mbps and for real video trace, we observe that the stalling ratio increases with increasing $\weight_{c}$ for CPH and CPH-EQ and later stabilizes. However, the buffer stalls are noticeably lower compared to CLIENT and CLIENT-CACHE.
The best setting for $\weight_{c}$ is 1.5 for this scenario as the cache hit ratio increases to approximately 12\% for CPH variants from 7\% when $\weight_{c}=1$.
If minimizing buffer stalls is more paramount for a WiFi network with a bottleneck link, then selecting $\weight_c=1$ provides the best trade-off between the cache hits and the stalling ratio. 
 
 \begin{figure*}[tb]
	\centering
	\subfloat[Video quality in Kbps.]{\includegraphics[width=0.33\linewidth]{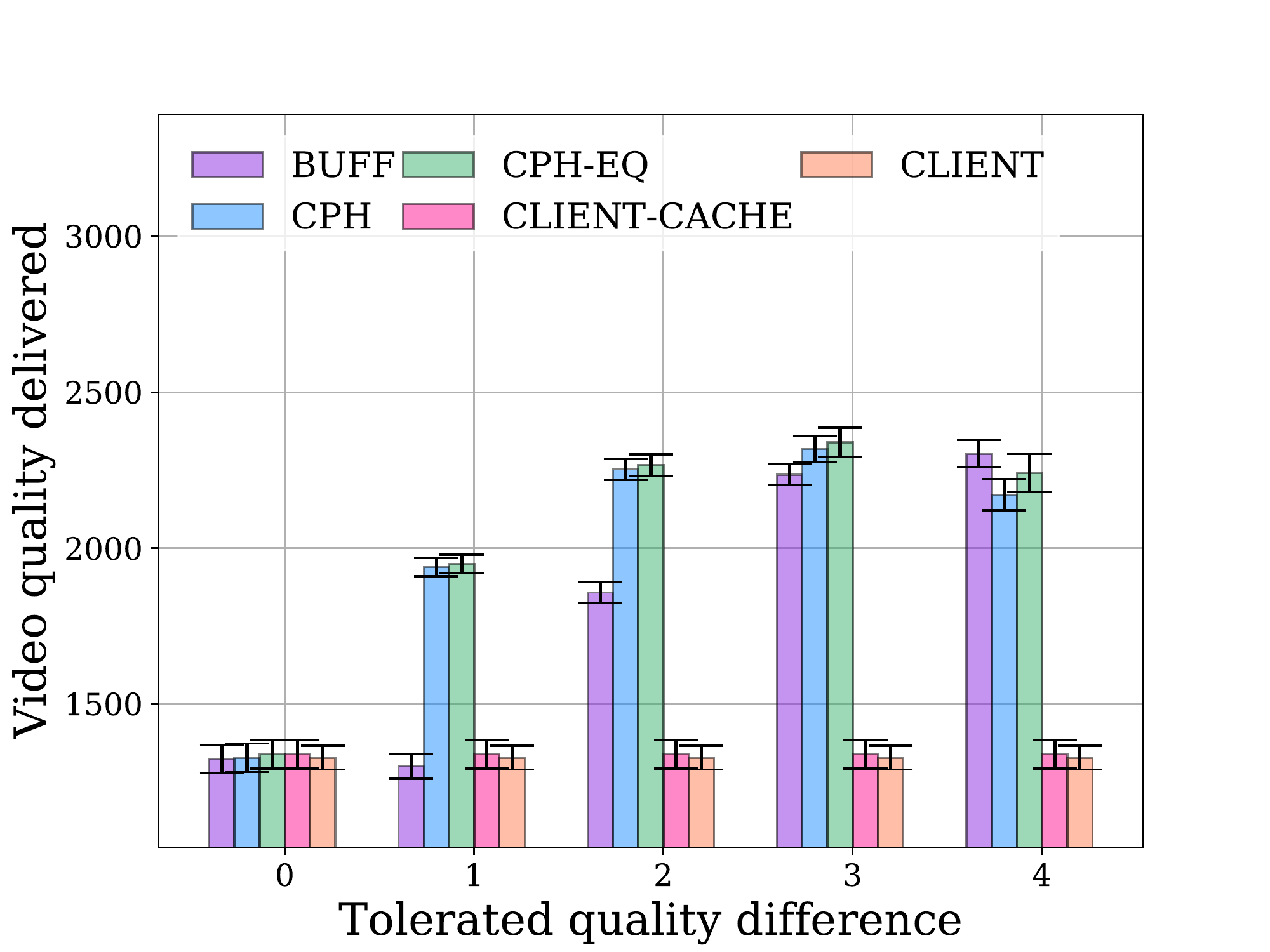}\label{fig:tol-videoquality}}
	\subfloat[Cache bit hit ratio.]{\includegraphics[width=0.33\linewidth]{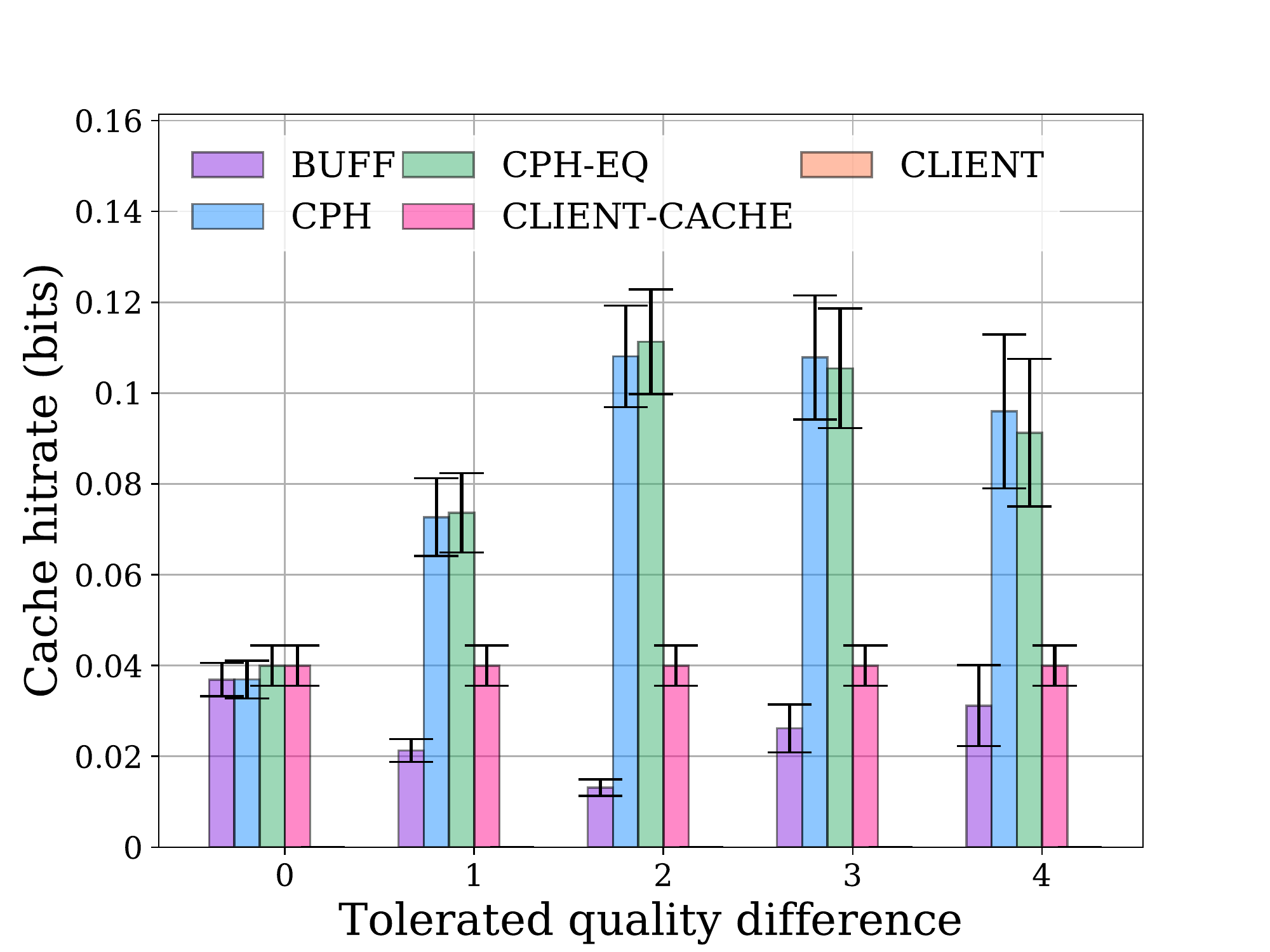}\label{fig:tol-cachehit}} 
	\subfloat[Stalling ratio.]{\includegraphics[width=0.34\linewidth]{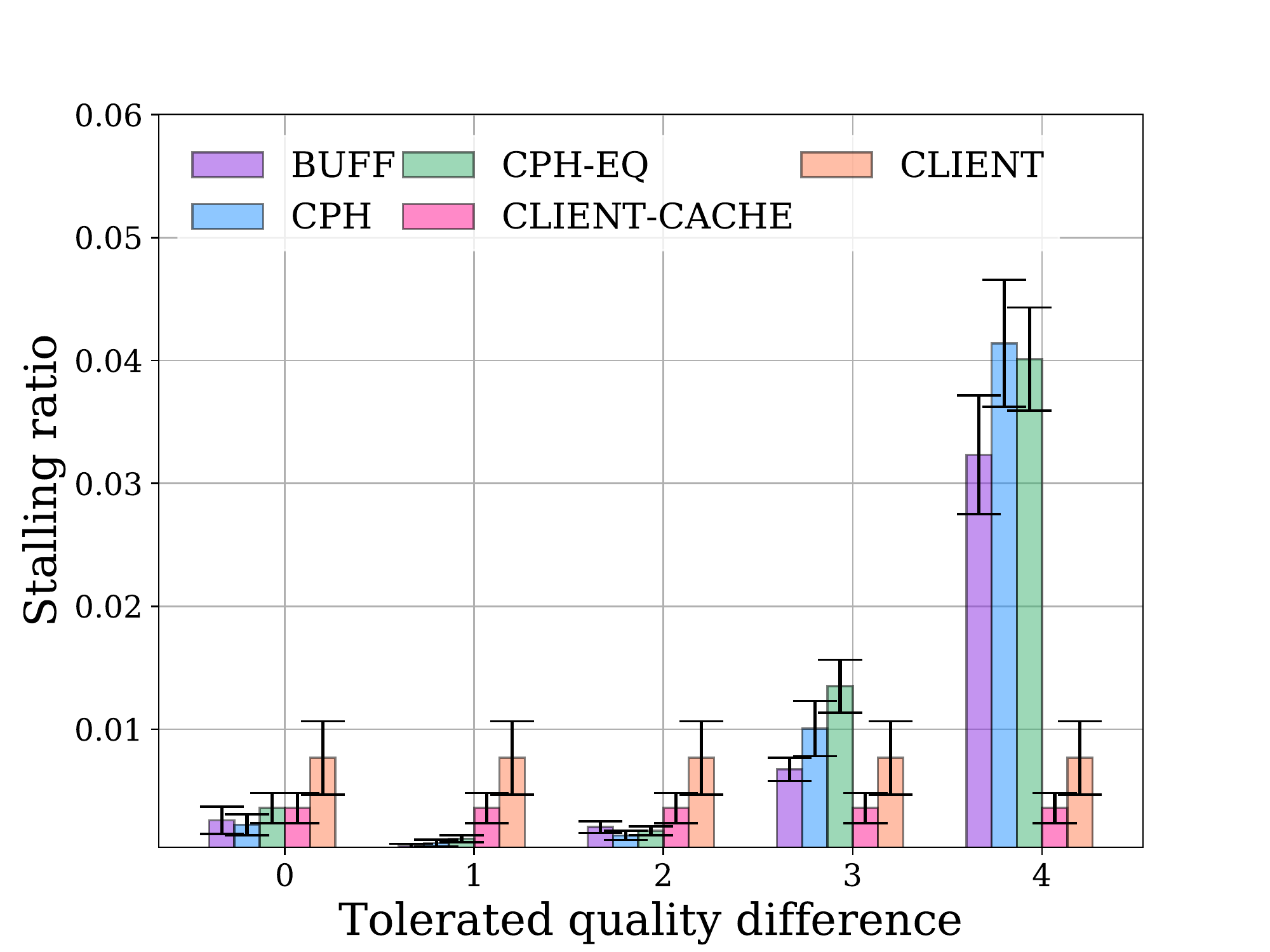}\label{fig:tol-Stallingratio}} 
	\caption{Impact of tolerated quality difference, 
	10 synthetic traces, 20 Mbps backhaul capacity. \label{fig:tol}}
	\vspace{-10pt}
\end{figure*}

\subsection{Impact of the tolerated quality difference}
Fig. \ref{fig:tol} shows the impact of the tolerated quality difference~($\toleratedQualityDiff$). 
Note that $\toleratedQualityDiff=0$ corresponds to the case where AP does not overwrite the client's decision. In other words, the AP serves the client only with the requested representation, which is naturally fetched from the cache upon availability. 
For $\toleratedQualityDiff=0$, our proposed schemes offer benefits compared to CLIENT in terms of cache hits~(Fig.\ref{fig:tol-cachehit})
in case of a relatively high backhaul capacity as in this case. When the backhaul capacity is a bottleneck, e.g., $\bottleneckCapacity=8$ Mbps, we observe also performance improvement in terms of video quality, stalling ratio, and initial latency~(not plotted).  
When $\toleratedQualityDiff$ is 1 or 2, we observe a significant improvement in cache hits in Fig. \ref{fig:tol-cachehit} and decrease in stalling ratio in Fig. \ref{fig:tol-Stallingratio} when our proposed schemes are used.  
If a significant change in client's decision is not desirable, $\toleratedQualityDiff$ can be set to 1 and CPH-variants can still achieve a noticeable improvement in cache hits. 
When $\toleratedQualityDiff=1$ and $\bottleneckCapacity=20$\,Mbps, CPH outperforms others. Under a lower capacity backhaul, CPH similarly achieves a higher video quality and cache hits but at the expense of increased buffer stalls compared to BUFF. Thus, depending on the primary goal~(higher cache hits or lower stalls), the AP can use BUFF or CPH in such cases.
\subsection{Impact of number of videos}
\begin{figure*}[tb]
	\centering
	\subfloat[Cache bit hit ratio.]{\includegraphics[width=0.33\linewidth]{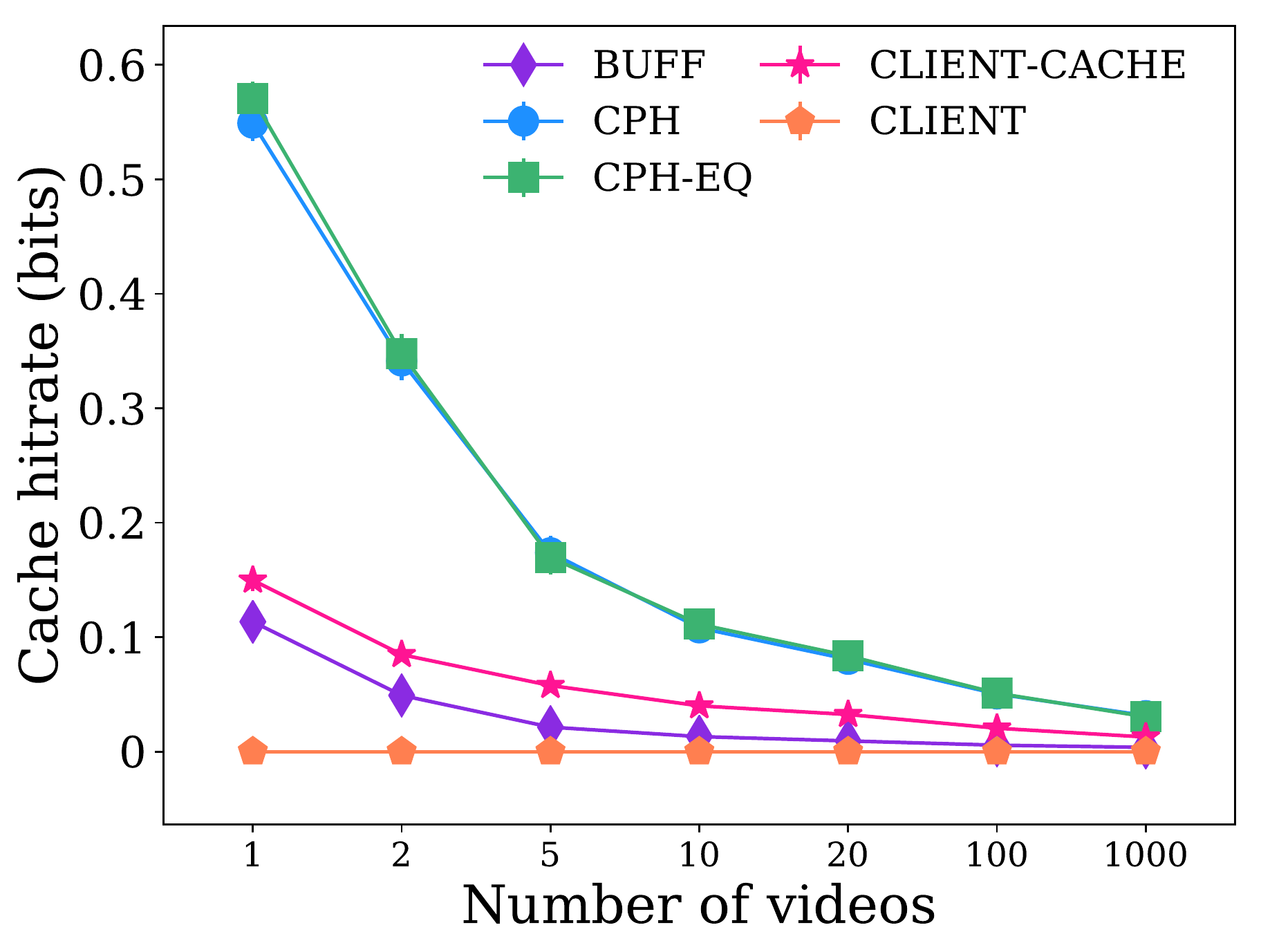}\label{fig:nvideo-cachehit}} 
	\subfloat[Video quality in Kbps.]{\includegraphics[width=0.33\linewidth]{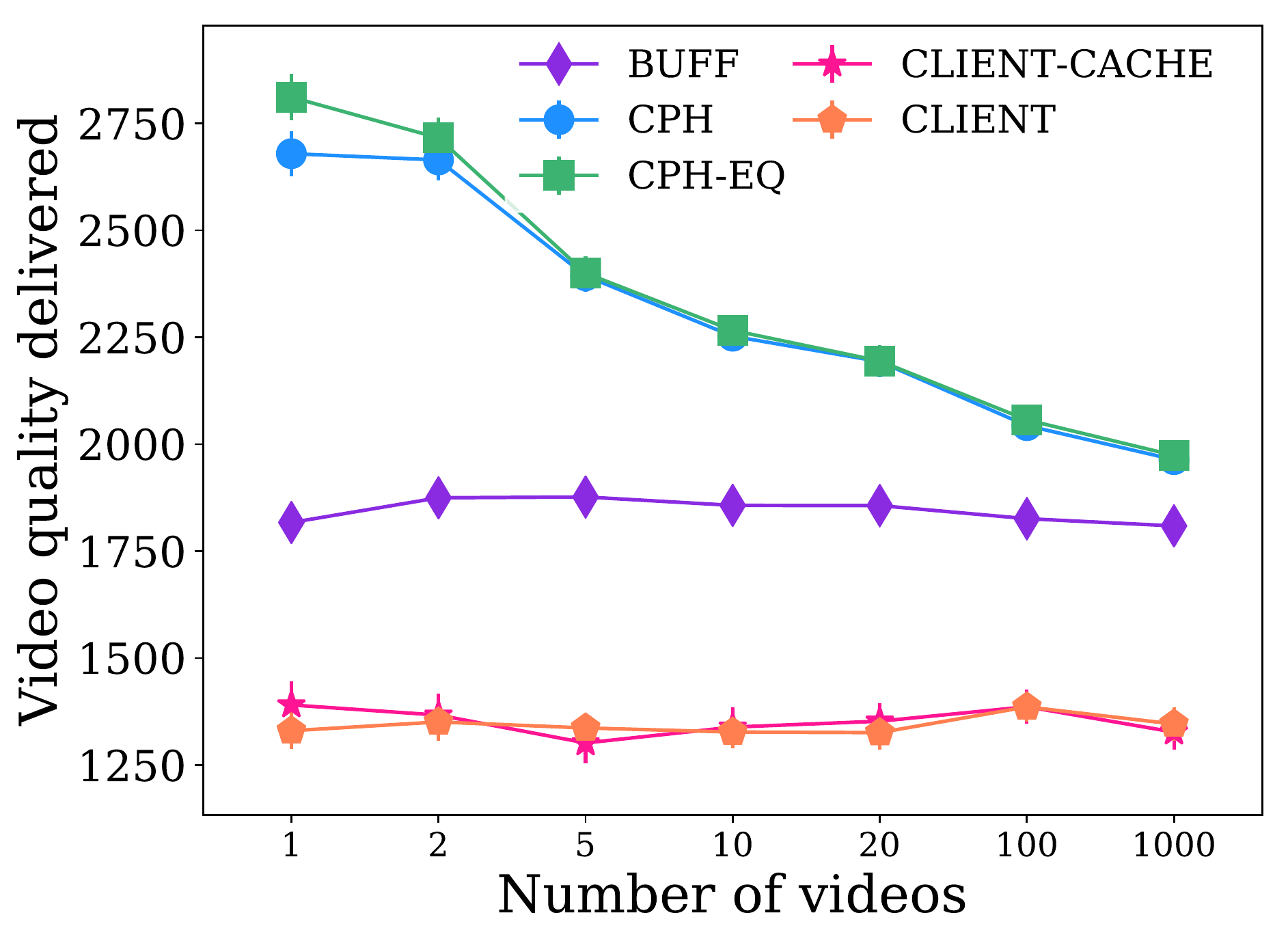}\label{fig:nvideo-videoquality}} 
	\subfloat[Backhaul utilization.]{\includegraphics[width=0.33\linewidth]{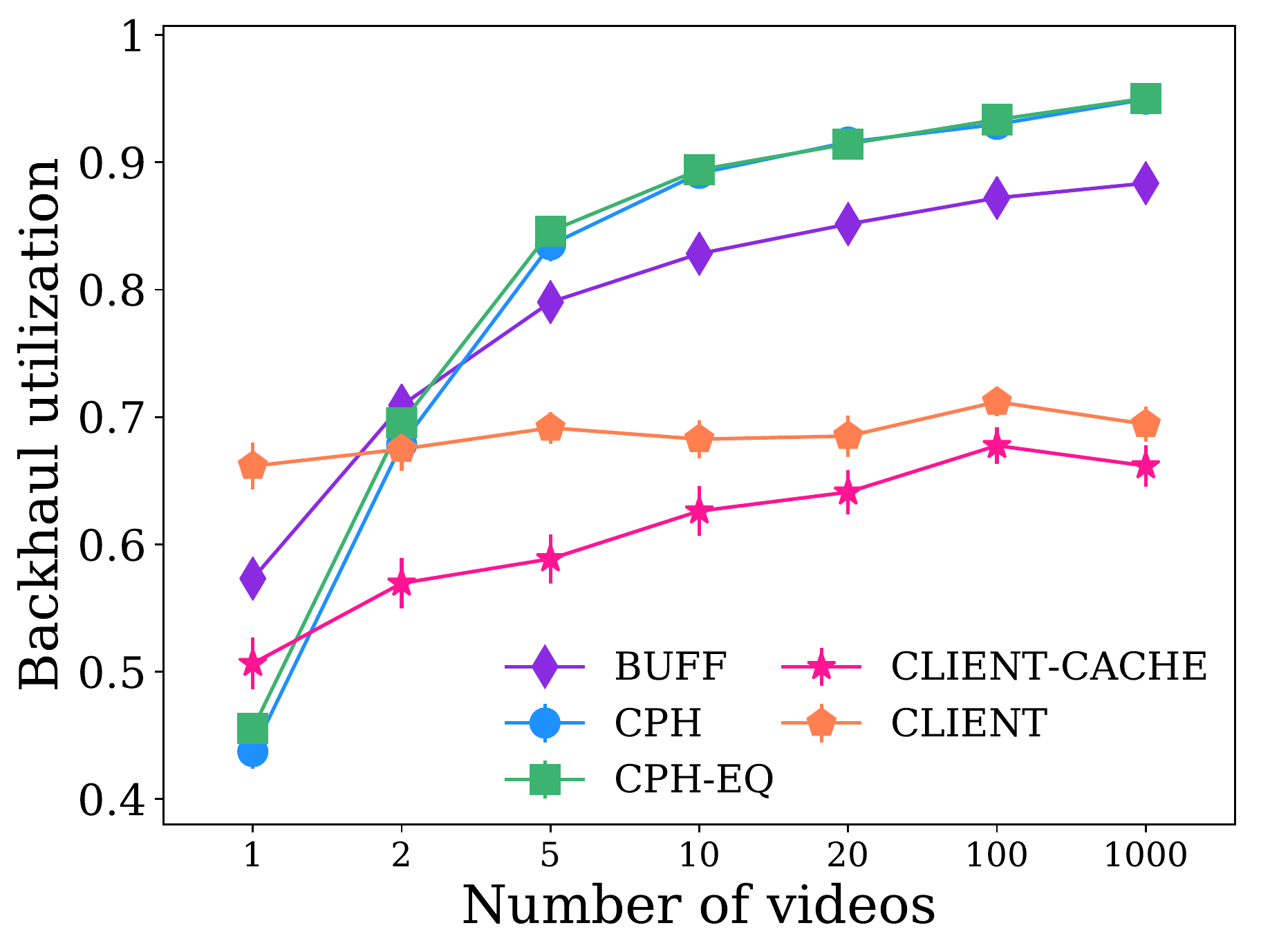}\label{fig:nvideo-bh}}
	\caption{Impact of number of videos, synthetic video data, $\bottleneckCapacity=$ 20 Mbps. \label{fig:nvideo}}
	\vspace{-10pt}
\end{figure*}

Fig. \ref{fig:nvideo} shows the change of performance with increasing number of video contents. If the traffic is only for one video, our schemes, which exploit the cache, provide a significant improvement in the cache hits~(Fig.~\ref{fig:nvideo-cachehit}). More precisely, when $\numVideo=1$, almost 57\% of the bits are served from the cache under CPH and CPH-EQ, compared to 15\% under CLIENT-CACHE and 11\% under BUFF. 
Delivering from the cache results in the client's rate adaptation algorithm to request a higher quality video. As a result, compared to BUFF and CLIENT variants, CPH and CPH-EQ offer better quality, which can be observed in Fig.~\ref{fig:nvideo-videoquality}.  
With more diverse contents, cache hit ratio decreases. 
\newtexthighlight{For a larger video catalog in the order of thousands~(e.g., the movie catalog of NetFlix\footnote{https://www.statista.com/statistics/563381/netflix-available-movies-by-country-in-europe/}), cache hits would indeed be lower. However, please recall that the simulations consider a limited number of requests spanning a period of approximately 10 minutes. In a practical setting with many more requests, the cache bit hitrate is expected to increase.
Nevertheless, our proposed schemes can still offer performance improvements over CLIENT in terms of video bitrates as shown in Fig. \ref{fig:nvideo-videoquality} and better utilization of the backhaul link as shown in Fig.~\ref{fig:nvideo-bh}. When $\bottleneckCapacity=8$ Mbps, we also observe improvement in stall ratio. This performance improvement is due to the AP's resource management approach that takes the clients' states, e.g., buffer levels, into account.}
Comparing BUFF and CPH variants, we observe the same trend as in the earlier scenarios: BUFF suffers from lower cache hits and video quality, but at the same time offers fewer buffer stalls when backhaul has limited capacity~(not plotted).

Note that our proposals might suffer from the sub-optimal decisions at the client's quality selection scheme as the AP considers the requested rate and deviate from it only within the limits of the tolerated quality level. Another option is to completely overwrite the client's requests which, however, conflicts with the client-driven nature of DASH.

\subsection{Discussion on a Practical System}
Here, we discuss briefly the implementation issues of \proposalName.
\newtexthighlight{An AP can calculate the utilities in (\ref{eq:utility}) using the following parameters: i) available bitrates of the requested video, ii) current client buffer level, iii) expected buffer level of a client after the delivery of the current chunk, and iv) minimum target buffer level. 
The AP and clients can use the following SAND messages encapsulated in HTTP header to convey this information:} \textit{AcceptedAlternatives}, \textit{DeliveredAlternative}, and \textit{ClientQoS}~\cite{sand2018}. Using AcceptedAlternatives, the client can notify the AP about the other quality levels it will accept. Using DeliveredAlternative, the AP can notify the client if it delivers an alternative rather than the requested quality. Using ClientQoS, the client can inform the AP about its buffer level. Using DeliveryBoostRequest, the client can request the network~(DANE) to assist it and ask for buffer boost. This message is defined in Network Assistance mode~\cite{sand2018} as an optional capability. Finally, DeliveryBoostResponse is sent as a response to DeliveryBoostRequest. The DANE can respond with this message that is defined in Network Assistance.

\newtexthighlight{A clear limitation of \proposalName~is that it is applicable only to HTTP traffic in which an AP can extract the contents of a video request message as necessary for the operation of  \proposalName. Given that an increasing fraction of Internet traffic is encrypted, it is important to design solutions that can work for encrypted content, as appears in  \cite{gutterman2019requet} and \cite{araldo2018caching} as examples. However, we leave this aspect to a future work since network assistance for encrypted content requires completely a new approach.}

\newtexthighlight{
Another possible direction is the design of cache management policies; 
a content-aware cache admission and replacement scheme can exploit the popularity of chunks and also use the information from clients, e.g., AnticipatedRequests. In our simulations, we intentionally considered a large cache to keep the impact of the cache admission and replacement policies minimal.}
Since the content providers such as YouTube collect user statistics, they might have a certain understanding of the popularity of each chunk and quality level. While such statistics are not publicly available, content providers would also benefit from sharing this information with network-assistance elements. 
Therefore, popularity values can be fed from the content provider to the network providers. 

\section{Conclusions \& Future Work}
\label{sec:conc}
Since video streaming is a dominant traffic accounting for a big share of network load, it is paramount to provide solutions to improve the performance of video traffic as well as to leverage some state-of-the-art approaches for decreasing the burden on the network. 
With this goal in mind, we propose network-side quality and resource assignment solutions running at a WiFi AP that takes advantage of the cached contents to both decrease the congestion in a bottleneck link and to 
improve the performance of video streaming clients, e.g., by offering higher video bitrate or lower buffer stalls.
Our simulations show the following: by a moderate adaptation of the clients' quality requests (e.g., offering a few quality levels more or less than requested levels), a WiFi AP can improve cache hits significantly while decreasing buffer stalls and increasing video bitrates. Moreover, the AP can allocate its downlink airtime considering the statistics of video clients such as buffer levels. While we have discussed how our proposals can be implemented using MPEG's SAND messages, as future work, we plan to provide a prototype and an evaluation of our schemes on this prototype using the state-of-the-art client driven DASH players to understand better the behavior of \proposalName~in interaction with TCP and real channel conditions.
%

\end{document}